\documentstyle[12pt]{article}
\def\simlt{\stackrel{<}{{}_\sim}}
\def\simgt{\stackrel{>}{{}_\sim}}
\def\be{\begin{equation}}
\def\ee{\end{equation}}
\def\bear{\be\begin{array}}
\def\eear{\end{array}\ee}
\def\bea{\begin{eqnarray}}
\def\eea{\end{eqnarray}}

\def\baselinestretch{1}
\begin{document}
\catcode`@=11
\newtoks\@stequation
\def\subequations{\refstepcounter{equation}%
\edef\@savedequation{\the\c@equation}%
  \@stequation=\expandafter{\theequation}
  \edef\@savedtheequation{\the\@stequation}
  \edef\oldtheequation{\theequation}%
  \setcounter{equation}{0}%
  \def\theequation{\oldtheequation\alph{equation}}}
\def\endsubequations{\setcounter{equation}{\@savedequation}%
  \@stequation=\expandafter{\@savedtheequation}%
  \edef\theequation{\the\@stequation}\global\@ignoretrue

\noindent}
\catcode`@=12
\begin{titlepage}
\title{{\bf Strong Constraints on the Parameter Space of the MSSM from
Charge and Color Breaking Minima}
\thanks{Research supported in part by: the CICYT, under
contracts AEN94-0928 (JAC) and AEN93-0673 (ALL, CM); the European Union,
under contracts CHRX-CT92-0004 (JAC), CHRX-CT93-0132 (CM) and
SC1-CT92-0792 (CM); the
Ministerio de Educaci\'on y Ciencia, under FPI grant (ALL).}
}
\author{ {\bf J.A. Casas${}^{ {\footnotesize\dag}}$},
{\bf A. Lleyda${}^{ {\footnotesize\ddag}}$ }
and {\bf C. Mu\~noz${}^{\footnotesize\ddag}$}\\
\hspace{3cm}\\
${}^{\footnotesize\dag}$ {\small Instituto de Estructura de la Materia CSIC}\\
{\small Serrano 123, 28006 Madrid, Spain}\\ {\small casas@cc.csic.es}\\
\vspace{-0.3cm}\\
${}^{\footnotesize\ddag}$ {\small Departamento de F\'{\i}sica
Te\'orica C--XI} \\
{\small Universidad Aut\'onoma de Madrid, 28049 Madrid, Spain}\\
{\small amanda@delta.ft.uam.es $$ cmunoz@ccuam3.sdi.uam.es}}
\date{}
\maketitle
\def\baselinestretch{1.15}
\begin{abstract}
\noindent
A complete analysis of all the potentially dangerous directions in the
field-space of the minimal supersymmetric standard model is carried out.
They are of two types, the ones associated with the existence of charge
and color breaking minima in the potential deeper than the
realistic minimum and the directions in the field-space along which the
potential becomes unbounded from below. The corresponding new constraints
on the parameter
space are given in an analytic form, representing a set of necessary and
sufficient conditions to avoid dangerous directions. They are very strong
and, in fact,
there are
extensive regions in the parameter space that become forbidden.
This produces important bounds, not only on the value of $A$, but also on
the values of $B$ and $M_{1/2}$. Finally, the crucial issue of the
one-loop corrections to the scalar potential has been taken into account in a
proper way.

\end{abstract}

\thispagestyle{empty}

\leftline{}
\leftline{}
\leftline{FTUAM 95/11}
\leftline{IEM--FT--100/95}
\leftline{June 1995}
\leftline{}

\vskip-23.cm
\rightline{}
\rightline{ FTUAM 95/11}
\rightline{ IEM--FT--100/95}
\vskip3in

\end{titlepage}
\newpage
\setcounter{page}{1}

\section{Introduction}

As is well known, the presence of scalar fields with
color and electric charge in supersymmetric (SUSY) theories induces the
possible existence of dangerous charge and color breaking (CCB)
minima, which would make the standard vacuum unstable. This is not
necessarily a shortcoming since many SUSY models can be discarded on
these grounds, thus improving the predictive power of the theory.

This fact has been known since the early 80's
\cite{Frere,Claudson}. Since then, several interesting papers have
appeared in the subject \cite{Drees,Gunion,Komatsu,Gamberini}.
However,
a complete study of this crucial issue is still lacking. This is
mainly due to two reasons. First, the enormous complexity of the
scalar potential, $V$, in a SUSY theory, which has motivated that
only analyses examining particular directions in the field--space
have been performed. Second, as we will see, the radiative
corrections to $V$ have not been normally included in a proper way.

Concerning the first point, and to introduce some notation, let us
write the tree-level scalar potential, $V_o$, in the minimal
supersymmetric standard model (MSSM):
\be
\label{Vo}
V_o = V_F + V_D + V_{\rm soft}\;\; ,
\ee
with
\subequations{
\be
\label{VF}
V_F = \sum_\alpha \left| \frac{\partial W}{\partial \phi_\alpha}
\right| ^2\;\;,
\ee
\be
\label{VD}
V_D = \frac{1}{2}\sum_a g_a^2\left(\sum_\alpha\phi_\alpha^\dagger
T^a \phi_\alpha\right)^2\;\; ,
\ee
\bea
\label{Vsoft}
V_{\rm soft}&=&\sum_\alpha m_{\phi_\alpha}^2
|\phi_\alpha|^2\ +\ \sum_{i\equiv generations}\left\{
A_{u_i}\lambda_{u_i}Q_i H_2 u_i + A_{d_i}\lambda_{d_i} Q_i H_1 d_i
\right.
\nonumber \\
&+& \left. A_{e_i}\lambda_{e_i}L_i H_1 {e_i} + {\rm h.c.} \right\}
+ \left( B\mu H_1 H_2 + {\rm h.c.}\right)\;\; ,
\eea}
\endsubequations
where $W$ is the MSSM superpotential
\be
\label{W}
W=\sum_{i\equiv generations}\left\{
\lambda_{u_i}Q_i H_2 u_i + \lambda_{d_i}Q_i H_1 d_i
+ \lambda_{e_i} L_i H_1 e_i \right\} +  \mu H_1 H_2\;\; ,
\ee
$\phi_\alpha$ runs over all the scalar components of the chiral superfields
and $a, i$ are gauge group and generation indices respectively.
$Q_i$ ($L_i$) are
the scalar partners of the quark (lepton) $SU(2)_L$ doublets and
$u_i,d_i$ ($e_i$) are the scalar partners of the quark (lepton)
$SU(2)_L$ singlets. In our notation $Q_i\equiv (u_L,\ d_L)_i$,
$L_i\equiv (\nu_L,\ e_L)_i$,
$u_i\equiv {u_R}_i$, $d_i\equiv {d_R}_i$, $e_i\equiv {e_R}_i$. Finally,
$H_{1,2}$
are the two SUSY Higgs doublets. The previous potential is extremely
involved since it  has a large number of independent fields.
Furthermore, even assuming universality of the soft breaking terms at
the unification scale, $M_X$, it contains a large number of
independent parameters: $m$, $M$, $A$, $B$, $\mu$, i.e. the universal
scalar and gaugino masses, the universal
coefficients of the trilinear and bilinear scalar terms, and
the Higgs mixing mass, respectively. In addition, there are the
gauge ($g$)
and Yukawa ($\lambda$)
couplings which are constrained by the experimental data. Notice that
$M$ does not appear explicitely in $V_o$, but it does through the
renormalization group equations (RGEs) of all the remaining parameters.

As mentioned above, the complexity of $V$ has made that only
particular directions in the field-space have been explored. It will
be useful for us to remind here two of them. First, there is the
``traditional" bound, first studied by Frere et al. and subsequently by
others \cite{Frere,Claudson}.
These authors considered just the three fields present in a
particular trilinear scalar coupling, e.g. $\lambda_u A_u Q_u H_2 u$, assuming
equal vacuum expectation values (VEVs) for them:
\be
\label{frerevevs}
|Q_u| = |H_2| = |u| \;\;,
\ee
where only the $u_L$-component of $Q_u$ takes a VEV in order to cancel the
D--terms. The phases of the three fields are taken in such way that the
trilinear scalar term in the potential has negative sign.
Then, they showed that a very deep
CCB minimum appears {\em unless} the famous constraint
\be
\label{frerebound}
|A_u|^2 \leq 3\left( m_{Q_u}^2 + m_{u}^2 + m_2^2\right)
\ee
is satisfied. In the previous equation $m_{Q_u}^2, m_{u}^2, m_2^2$ are the
mass parameters of $Q_u$, $u$, $H_2$. Notice from eq.(\ref{Vo}) that
$m_2^2$ is the sum of the $H_2$ squared soft mass, $m_{H_2}^2$, plus
$\mu^2$. Similar
constraints for the other trilinear terms can straightforwardly be
written.  These ``traditional" bounds have extensively been used in
the literature.
The second example is due to Komatsu \cite{Komatsu}, who
realized that the potential of eq.(\ref{Vo}) along the direction
\bea
\label{komatsuvevs}
&|L_i|^2& =|H_2|^2 +
|Q_j|^2\nonumber \\
&Q_jd_j& = -
\frac{\mu}{\lambda_{d_j}} H_2
\nonumber \\
&|Q_j|^2& =|d_j|^2
 \;\;,
\eea
with $L_i$ and $Q_j$ VEVs taken along $\nu_L$ and $d_L$
respectively, is unbounded from
below (UFB) {\em unless} the constraint
\be
\label{komatsubound}
m_2^2 - \mu^2 + m_{L_i}^2 \geq 0
\ee
is satisfied. Komatsu claimed that for $M_{\rm top}=100$ GeV this constraint
is extremely strong. To see this, notice that at the $M_Z$ scale
$m_2^2 - \mu^2$ is normally negative and of the same order as $m_{L_i}^2$.

Let us go now to the issue of the radiative corrections. Usually, the
scalar potential is considered at tree-level, improved by
one-loop RGEs, so that all the parameters appearing in it (see eq.(\ref{Vo}))
are running
with the renormalization scale, $Q$. Then it is demanded that the
previous CCB constraints, i.e. eqs.(\ref{frerebound}),
(\ref{komatsubound}) and others, are satisfied at any scale between
$M_X$ and $M_Z$. However, as was clarified by Gamberini et al.
\cite{Gamberini}, this is not correct. $V_o$ is strongly $Q$--dependent
and the one-loop radiative corrections to it, namely
\be
\label{V1}
\Delta V_1={\displaystyle\sum_{\alpha}}{\displaystyle\frac{n_\alpha}{64\pi^2}}
M_\alpha^4\left[\log{\displaystyle\frac{M_\alpha^2}{Q^2}}
-\frac{3}{2}\right]\;\;,
\ee
are crucial to make the potential stable against variations of the $Q$
scale. In eq.(\ref{V1}) $M_\alpha^2(Q)$ are the improved tree-level
(field--dependent) squared mass eigenstates and
$n_\alpha=(-1)^{2s_\alpha} (2s_\alpha+1)$, where $s_\alpha$
is the spin of the corresponding
particle. Clearly, the complete one-loop potential $V_1=V_o+\Delta V_1$
has a structure that is even far more involved than $V_o$ (notice that
$\Delta V_1$ is a complicated function of all the scalar fields). This makes in
practice the minimization of the complete $V_1$ an impossible task. However,
in the region of $Q$ where $\Delta V_1$ is
small, the predictions of $V_o$ and $V_1$ essentially coincide. This
occurs for a value of $Q$ of the order of the most significant $M_\alpha$
mass appearing in (\ref{V1}), which in turn depends on what is the
direction in the field-space that is
being analyzed. Moreover, this corresponds to the
region of maximal $Q$--invariance of $V_1$
\cite{Gamberini,CC}. Therefore,
one can still work just with $V_o$, but with the appropriate choice
of $Q$.
In this way it was shown in ref.\cite{Gamberini}
that the apparently very strong constraint (\ref{komatsubound}) was
in fact extremely weak. It should be mentioned however that the
analysis was performed assuming $M_{\rm top}=M_W$. As we will see
in sect.3 and sect.6,
once the constraint (\ref{komatsubound}) is improved and the top
quark mass is set at its current value, the corresponding bound is
really very restrictive.

\vspace{0.3cm}
\noindent To summarize the situation, due to the complexity of the
SUSY scalar potential, only particular directions in the
field-space have been considered, thus obtaining necessary but not
sufficient conditions to avoid dangerous CCB minima. Furthermore,
the usual lack of an optimum scale to evaluate the constraints
implies that their restrictive power has been normally overestimated.
E.g., eq.(\ref{frerebound})-type constraints when (incorrectly) analyzed
at $M_X$ are very strong.
The aim of this paper is to improve, and hopefully fix, this
situation.

\vspace{0.3cm}
\noindent
In sect.2 we review the realistic minimum that corresponds to the standard
vacuum.
In particular, we derive the correct scale at which the minimization of the
potential has to be evaluated and summarize all the theoretical
and experimental constraints that the realistic minimum must
satisfy. In sect.3 we carry out a complete analysis of all the potentially
dangerous directions in the field-space along which the potential can
become unbounded from below, obtaining the corresponding constraints on
the parameter space. The possibility of spontaneous lepton number breaking is
also discussed since one of those directions involves the sneutrino.
In sect.4 we perform a complete analysis of all the constraints arising from
the
existence of charge and color breaking minima in the potential deeper
than the realistic minimum.
Let us remark that the bounds obtained in this section, as well as in sect.3,
are completely general and  are expressed in an analytical way. Hence, they
represent necessary and sufficient conditions on the parameters of the
MSSM, which can also be applied to the non-universal case. The
correct choice of the scale to evaluate the constraints is also discussed.
The reader not interested
in the precise details of the calculation of the constraints
may jump over the
two previous
sections and go directly to sect.5, where we summarize all the previous
results. In sect.6 we analize numerically how the previously found constraints
restrict the whole parameter space of the MSSM.
Although the ``traditional" bounds evaluated
at the correct scale turn out to be very weak, we will show that
the new charge and color breaking constraints found here
are much more important and,
in fact, there are extensive regions in the parameter space which are
forbidden. The unbounded from below-like constraints
turn out to be even stronger.
All together produces important bounds not only on the value of
$A$, but also on the values of $B$ and $M$.
The conclusions
are left for sect.7. Finally, the Appendix is devoted to the proof of some
relevant general properties concerning CCB minima which are used throughout
the paper.

\section{The realistic minimum}

The neutral part of the Higgs potential in the MSSM is
\bea
\label{Vhiggs}
V_{\rm Higgs}&=&m_1^2 |H_1|^2 + m_2^2 |H_2|^2 - 2|m_3^2||H_1||H_2|
+ \frac{1}{8}(g'^2+g_2^2)(|H_2|^2-|H_1|^2)^2
\nonumber \\
&+& \Delta V_1\;\;,
\eea
where $m_1^2\equiv m_{H_1}^2 + \mu^2$, $m_2^2\equiv m_{H_2}^2 + \mu^2$,
$m_3^2\equiv-\mu B$,
$g_3=g_2=g_1=\sqrt{\frac{5}{3}}g'$ at $M_X$, and $\Delta V_1$ is given in
eq.(\ref{V1}). It should develope a minimum at
$|H_1|=v_1$, $|H_2|=v_2$, such that
$v_1^2+v_2^2=2M_W^2 / g_2^2$. This is the realistic minimum
that corresponds to the standard vacuum.
In this way the requirement of correct electroweak breaking fixes one
of the five independent parameters of the MSSM (i.e. $m,M,A,B,\mu$),
say $\mu$.
Actually, for some choices of the four remaining parameters
($m,M,A,B$), there is no value of $\mu$ capable of producing the
correct electroweak breaking. Therefore, this requirement restricts
the parameter space further, as is illustrated in Fig.1 (central darked
region) with a representative example.
The value of the potential at the realistic minimum is
\be
\label{Vreal}
V_{\rm real\;min}
=- \frac{1}{8} \;(g'^2+g_2^2)\;(v_2^2-v_1^2)^2 \\
= - \frac{ \left \{ \; \left[ \;( m_1^2+m_2^2 )^2-4 |m_3|^4 \; \right]
^{1/2}  - m_1^2+m_2^2 \; \right \} ^2  } {2 \; (g'^2+g_2^2) } \;\; .
\ee
Note that this is the result obtained by minimizing
just the tree-level part of (\ref{Vhiggs}).
As explained in sect.1 this is correct if the minimization
is performed at some sensible scale around
which $V_{\rm Higgs}$ is $Q$-invariant. We have chosen for this the
scale $Q=M_S$, where the predictions for $v_{1,2}$ from $V_{\rm Higgs}$
with and without radiative corrections coincide\footnote{Strictly,
this can only be demanded for one of the two Higgs VEVs, say $v_2$,
but then it also occurs for $v_1$ with high accuracy.}.
More precisely, the requirement
$\left.\frac{\partial \Delta V_1}{\partial H_2}\right|_{Q=M_S}=0$
gives
\bea
\label{MS}
M_S&=&e^{-1/2} \; \prod_{\alpha} M^
{\frac{d_\alpha M^2_\alpha}{ \sum_\beta d_\beta M^2_\beta}}_\alpha\\
d_\beta&=&n_\beta \frac{\partial M^2_\beta}{\partial  H_2}\;\; .
\eea
Note that $M_S$ is a certain average of typical SUSY masses.

In all the previous calculation, one has to run the parameters
through their respective RGEs, which depend on the value of the gauge
and Yukawa couplings. The boundary conditions for these are
determined by the experimental values of $\alpha_1(M_Z)$,
$\alpha_2(M_Z)$, $\alpha_3(M_Z)$ and the quark masses. In particular,
we take $M^{\rm phys}_{\rm top}=174$ GeV as the physical (pole) top
mass, which is related to the running top mass through a standard
expression \cite{Arason}. Actually, not for all the parameter space
it is possible to choose the boundary condition of $\lambda_{top}$
so that the experimental mass is reproduced because the RG infrared
fixed point of $\lambda_{top}$ puts an upper bound on $M_{\rm top}$,
namely $M_{\rm top}\simlt 197\sin\beta$ GeV \cite{infrared},
where $\tan\beta=v_2/v_1$.
The corresponding restriction in the parameter space is certainly
substantial as is illustrated in Fig.1 (upper and lower darked regions).
Let us also mention that whenever $\tan \beta$ is not large
($\simlt 10$), it is a good approximation to neglect the effect
of the bottom and tau Yukawa couplings in the set of RGEs. We have
adopted this simplification throughout the paper.

To be considered as realistic, the previous minimum has to satisfy
a number of further constraints. First of all, $V_{\rm Higgs}$ should
not be unbounded from below. Working just with the tree-level part of
(\ref{Vhiggs}), this leads to the well-known condition
\be
\label{ufbhiggs}
m_1^2+m_2^2 \ge 2|m_3^2|\;\;.
\ee
Actually, (\ref{ufbhiggs}) is automatically satisfied at $Q=M_S$, but
this is not necessarily true for $Q>M_S$. If it is not, then for
large VEVs of the Higgs fields ($ H_{1,2}\sim Q>M_S$),
the potential becomes much deeper than the realistic minimum. Hence,
we must impose (\ref{ufbhiggs}) at any $Q>M_S$ and, in particular, at
$Q=M_X$. Very often the additional condition
\be
\label{existmin}
m_1^2m_2^2-|m_3|^4 < 0\;\;,
\ee
is demanded at the $M_S$ scale to ensure that the $H_1=H_2=0$
(non-electroweak-breaking) point is unstable. However, it can
be checked that (\ref{existmin}) is automatically satisfied
once a realistic minimum has been found.

Second, we must be sure that the realistic minimum of the (neutral)
Higgs potential is really a
minimum in the whole field-space. This simply implies that all the
scalar squared mass eigenvalues (charged Higgses, squarks and
sleptons) must be positive. This is guaranteed for the charged Higgs
fields since in the MSSM the minimum of the Higgs potential always
lies at
\be
\label{chargedhiggs}
H_2^+ = H_1^- = 0\;\;,
\ee
but not for the rest of the sparticles.
Actually, we have verified that the charged Higgs fields do not play
any significant role not only for the realistic minimum, but also for
any CCB direction. So, we have assumed (\ref{chargedhiggs})
throughout the paper.
Finally, we must go further and demand that all the not yet
observed particles, i.e.
gluino ($g$), charginos ($\chi^\pm$), neutralinos
($\chi^o$), Higgses, squarks ($q$)
and sleptons ($l$),
have masses compatible with the experimental bounds. Conservatively enough,
we have imposed
\bea
\label{Expb}
M_{g}&\geq&120\ {\rm GeV}\;,\; \; M_{\chi^{\pm}}\geq 45\ {\rm GeV}
\nonumber \\
M_{\chi^o}&\geq&18\ {\rm GeV}\;,\; \; M_{q}\geq 100\ {\rm GeV}
\nonumber \\
M_{t}&\geq&45\ {\rm GeV}\;,\; \; M_{l}\geq 45\ {\rm GeV}
\;\; ,
\eea
in an obvious notation. The effect of strengthening these bounds can
be trivially incorporated to the results of the paper.

\section{Improved UFB constraints}

These constraints arise from directions in the field-space along
which the (tree-level) potential can become unbounded from below (UFB).
It is interesting to note that usually this is only true at tree-level
since radiative corrections eventually raise the potential for
large enough values of the fields. This is the case of UFB-2,3 directions
studied below. We have already mentioned the
UFB direction of eq.(\ref{komatsuvevs}) \cite{Komatsu}, and the one
in the Higgs part of the potential involving only the Higgs
fields (see eq.(\ref{ufbhiggs})). However, as we are about to see, it
is possible to do a complete clasification
of {\em all} the potentially dangerous UFB directions and constraints in the
MSSM.
We will also consider the radiative corrections in a proper way by making
an suitable choice of the renormalization scale (for more details see
subsect.4.5).

\subsection{General properties}

\begin{enumerate}

\item[{\bf 1}]
It is easy to check that trilinear scalar terms cannot play a
significant role along an UFB direction since for large enough values
of the fields the corresponding quartic (and positive) F--terms
become unavoidably larger.

\item[{\bf 2}]
Since all the physical masses must be positive at $Q=M_S$,
the only negative terms in the (tree-level)
potential that can play a relevant role along an UFB direction
are\footnote{The only possible exception are the stop soft mass terms
$m_{Q_t}^2 |Q_t|^2+m_{t}^2 |t|^2$ since the stop masses are given by
$\sim (m_{Q_t,t}^2 +M_{top}^2 \pm \;{\rm mixing})$, but this possibility
is barely consistent with the present bounds on squark masses.}
\be
\label{ufbneg}
m_2^2 |H_2|^2\;,\;\;\;-2|m_3^2| |H_1||H_2|\;\;\;\;.
\ee
Therefore, any UFB direction must involve, $H_2$ and, perhaps,
$H_1$. Furthermore, since the previous terms are cuadratic, all the
quartic (positive) terms coming from F-- and D--terms must be
vanishing or kept under control along an UFB direction. This means
that, in any case, besides $H_2$ some additional field(s) are required.

\end{enumerate}

\subsection{UFB constraints}

Using the previous general properties we can completely clasify the
possible UFB directions in the MSSM:

\begin{description}

\item[UFB-1]
${}^{}$\\
The first possibility is to play just with $H_1$ and $H_2$. Then,
the relevant terms of the potential are those written
in eq.(\ref{Vhiggs}). Obviously, the only possible UFB direction
corresponds to choose $H_1=H_2$ (up to $O(m_i)$ differences
which are negligible for large enough values of the fields),
so that the quartic D--term is cancelled. Thus, the (tree-level) potential
along the UFB-1 direction is
\be
\label{Vnuevo}
V_{\rm UFB-1}=(m_1^2 + m_2^2 - 2|m_3^2|) |H_2|^2  \ .
\ee
The constraint to be imposed is that, for any value of $|H_2|<M_X$,
\be
\label{nuevo}
V_{\rm UFB-1}(Q=\hat Q)> V_{\rm real\;min}(Q=M_S) \ ,
\ee
where $V_{\rm real\;min}$ is the value of the realistic minimum,
given by eq.(\ref{Vreal}), and $V_{\rm UFB-1}$ is evaluated at an
appropriate scale $\hat Q$. (Recall that since
we are dealing with the tree-level part
of the Higgs potential, this has to be computed at a
correct renormalization scale.) More
precisely $\hat Q$ must be of the same order
as the most significant mass along this UFB-1 direction, which is
$\hat Q \sim {\rm Max}(g_2|H_2|,\ \lambda_{top}|H_2|,\ M_S)$.

However, in this case, as already discussed in sect.2, eq.(\ref{nuevo})
is accurately equivalent to the well-known condition
\be
\label{ufb1}
m_1^2+m_2^2 \ge 2|m_3^2|
\ee
evaluated at any $Q>M_S$ and, in particular, at $Q=M_X$.
If this is not satisfied the potential eq.(\ref{Vnuevo})
is always deeper than the realistic minimum.

\item[UFB-2]
${}^{}$\\
If we include some additional field (besides $H_2,H_1$), this can
only be justified in order to cancel (or keep under control) the
D--terms in a more efficient way than just with $H_1$. It is easy to
see by simple inspection that the best possible choice is a slepton
$L_i$ (along the $\nu_L$ direction), since it has the
lightest mass without contributing to further quartic terms in $V$.
Consequently, the relevant potential reads
\be
\label{Vrel}
V=m_1^2 |H_1|^2 + m_2^2 |H_2|^2 - 2|m_3^2||H_1||H_2|
+ m_{L_i}^2 |L_i|^2 +
\frac{1}{8}(g'^2+g_2^2)( |H_2|^2-|H_1|^2-|L_i|^2 )^2.
\ee
It is now straightforward to write the deepest direction along
the $L_i$, $H_1$ variables, namely\footnote{
It is trivial to check that the remaining condition in order to get
a true minimum in the tree-level potential of eq.(\ref{Vrel}), $\partial
V/\partial H_2=0$, cannot be fulfilled.
This result contradicts the usual statement that can be found in the
literature, namely that (tree-level) spontaneous lepton number
breaking, and therefore R--parity breaking,
generating a majoron is possible in SUSY without
introducing additional fields, since the scalar partner of the neutrino
may acquire a non--vanishing VEV \cite{Valle}.}
\subequations{
\be
\label{L2ufb2}
|L_i|^2 = \frac{-4m_{L_i}^2}{g'^2+g_2^2}+|H_2|^2-|H_1|^2\;\;,
\ee
\be
\label{H1ufb2}
|H_1| = |H_2| \frac{|m_3^2|}{m_1^2-m_{L_i}^2}
=|H_2|\frac{|m_3^2|}{\mu^2}\;\;,
\ee}
\endsubequations
provided that
\subequations{
\be
\label{m3mu}
|m_3^2| < \mu^2
\ee
\be
\label{H2geq}
|H_2|^2 > \frac{ 4m_{L_i}^2 } { (g'^2+g_2^2)
        \left[ 1- \frac{|m_3|^4}{\mu^4} \right] } \;\;,
\ee}
\endsubequations
otherwise the optimum value for $L_i$ is $L_i=0$, and we come back to
the direction UFB-1. From (\ref{Vrel}),
(\ref{L2ufb2}), (\ref{H1ufb2})
we can write the (tree-level) potential along the UFB-2 direction
\be
\label{Vufb2}
V_{\rm UFB-2}=\left[m_2^2 + m_{L_i}^2 - \frac{|m_3|^4}{\mu^2}\right]
|H_2|^2 - \frac{2m_{L_i}^4}{g'^2+g_2^2}
\;\;.
\ee
 From (\ref{Vufb2}) it might seem that the potential is unbounded from
below unless
\be
\label{ufb2}
m_2^2+m_{L_i}^2-\frac{|m_3|^4}{\mu^2}\geq 0\;\;.
\ee
However, what should be really verified is that, for any value of
$|H_2|<M_X$ satisfying (\ref{H2geq}),
\be
\label{condufb2}
V_{\rm UFB-2}(Q=\hat Q)> V_{\rm real\;min}(Q=M_S)
\;\;,
\ee
where $V_{\rm real\;min}$ is the value of the realistic minimum,
given by eq.(\ref{Vreal}), and $V_{\rm UFB-2}$ is evaluated at an
appropriate scale $\hat Q$. More precisely $\hat Q$ must be of the same order
as the most significant mass along this UFB-2 direction, which is
$\hat Q \sim {\rm Max}(g_2|H_2|,\ \lambda_{top}|H_2|,\ M_S)$.

This direction
is dangerous not only because in general
the Higgses get too large VEVs but
also because the breaking of lepton number through the VEV of the
sneutrino leads to the existence of a majoron already excluded by
experimental results \cite{Valle2}.

Let us finally note that the last identity of eq.(\ref{H1ufb2}) relies
on the equality $m_1^2-m_{L_i}^2=\mu^2$, which only holds
under the assumption of degenerate soft scalar masses for
$H_1$ and $L_i$ at $M_X$ and
in the
approximation of neglecting the bottom and tau Yukawa couplings
in the RGEs. Otherwise, one simply must replace
$\mu^2$ by $m_1^2-m_{L_i}^2$ everywhere in eqs.(22--\ref{ufb2}).

\item[UFB-3]
${}^{}$\\
The only remaining possibility is to take $H_1=0$. Then, the $H_1$
F--term can be cancelled with the help of the VEVs of $d$--type
squarks of a particular generation, say $d_{L_j},d_{R_j}$,
without contributing to further quartic terms.
More precisely
\be
\label{trick}
\left|\frac{\partial W}{\partial H_1}\right|^2=
\left|\mu H_2 + \lambda_{d_j} d_{L_j} d_{R_j} \right|^2 = 0
\;\;.
\ee
Taking the VEVs $d_{L_j} = d_{R_j} \equiv d$, the SU(3) D--term
remains vanishing.
The main
consequence of taking these VEVs as in eq.(\ref{trick}) is to modify
the $H_2$ mass term from $m_2^2 |H_2|^2$ to $(m_2^2 - \mu^2)|H_2|^2$.
It is important to note that this trick cannot be used if $H_1\neq
0$, as happens in the UFB--2 direction, since then the $d_{L_j},d_{R_j}$
F--terms would
eventually dominate. Now, in order to cancel (or keep under control)
the $SU(2)_L$ and $U(1)_Y$ D--terms we need the VEV of some additional field,
which cannot be $H_1$ for the above mentioned reason. Once again the
optimum choice is a slepton $L_i$ along the $\nu_L$ direction, as in the
UFB--2 case.
Consequently, the relevant potential reads
\be
\label{Vrel3}
V=(m_2^2-\mu^2) |H_2|^2 + (m_{Q_j}^2+m_{d_j}^2) |d|^2 + m_{L_i}^2 |L_i|^2 +
\frac{1}{8}(g'^2+g_2^2)(|H_2|^2+|d|^2-|L_i|^2)^2.
\ee
This was the kind of possible UFB direction first
noticed in the interesting work of ref.\cite{Komatsu}
taking a particular combination
of the VEVs of $H_2,d_{L_j},d_{R_j},L_i$
(see eq.(\ref{komatsuvevs})), which
is not the optimum one. It is straightforward to see that
the deepest direction in the field--space is
\subequations{
\be
\label{L2ufb3}
|L_i|^2 = \frac{-4m_{L_i}^2}{g'^2+g_2^2}+(|H_2|^2+|d|^2)\;\;,
\ee
\be
\label{dufb3}
d^2 =- \frac{\mu}{\lambda_{d_j}} H_2
\;\;,
\ee}
\endsubequations
provided that
\be
\label{H2ufb3}
|H_2| >\sqrt{\frac{\mu^2}{4\lambda_{d_j}^2}+\frac{4m_{L_i}^2}{g'^2+g_2^2}}
-\frac{|\mu|}{2\lambda_{d_j}}\;\;,
\ee
otherwise the optimum value for $L_i$ is $L_i=0$. Now,
from (\ref{Vrel3}), (\ref{L2ufb3}), (\ref{dufb3}),
we can write down the (tree-level) potential along the UFB-3 direction
\be
\label{Vufb3}
V_{\rm UFB-3}=\left[m_2^2 -\mu^2+ m_{L_i}^2 \right]
|H_2|^2 + \frac{|\mu|}{\lambda_{d_j}}\left[m_{Q_j}^2 +m_{d_j}^2+ m_{L_i}^2
\right]
|H_2| - \frac{2m_{L_i}^4}{g'^2+g_2^2}
\;\;.
\ee
If $H_2$ does not satisfy (\ref{H2ufb3}), then
\be
\label{Vufb3p}
V_{\rm UFB-3}=\left[m_2^2 -\mu^2\right]
|H_2|^2 + \frac{|\mu|}{\lambda_{d_j}}\left[m_{Q_j}^2 +m_{d_j}^2\right]
|H_2| + \frac{1}{8}(g'^2+g_2^2)\left[ |H_2|^2+\frac{|\mu|}{\lambda_{d_j}}
|H_2|  \right]^2.
\ee
Analogously to the UFB--2 case, what should be demanded is that, for
any value of $|H_2|<M_X$,
\be
\label{condufb3}
V_{\rm UFB-3}(Q=\hat Q)> V_{\rm real\;min}(Q=M_S)
\;\;,
\ee
where $V_{\rm real\;min}$ is the value of the realistic minimum,
given by eq.(\ref{Vreal}), and $V_{\rm UFB-3}$ is evaluated at an
appropriate scale $\hat Q$. In this case
$\hat Q \sim {\rm Max}\left(g_3 |d|,\  \lambda_{u_j}|d|,\  g_2|H_2|,\right.$
$\left.\lambda_{top}|H_2|,\  g_2|L_i|,\  M_S\right)$. From
eqs.(\ref{Vufb3}--\ref{condufb3}),
it is clear that the larger $\lambda_{d_j}$ the more restrictive
the constraint becomes. Consequently, the optimum choice of
the $d$--type squark is the third generation one, i.e.
$d_j=$ sbottom. We have considered anyway the three possibilities,
confirming this expectative.

Finally, it is relevant to note that the job of the $d_{L_j}, d_{R_j}$
squarks in eq.(\ref{trick}) can be done by $e_{L_j}, e_{R_j}$ sleptons
with $j\neq i$ (this was not noted in ref.\cite{Komatsu}). Then everything
between
eq.(\ref{trick}) and eq.(\ref{condufb3}) remains
identical with the substitutions
\bea
\label{otramas}
d \rightarrow e\;,\hspace{2.0cm}
\lambda_{d_j} \rightarrow \lambda_{e_j}\;,\hspace{2.0cm}
Q_j \rightarrow L_j\;.
\eea
This is true in particular for eq.(\ref{condufb3}) and
eqs.(\ref{Vufb3},\ref{Vufb3p}), which represent the form of the UFB-3 bound.
The appropriate scale, $\hat Q$, to evaluate $V_{\rm UFB-3}$ is now given by
$\hat Q \sim {\rm Max}(g_2 |e|, g_2|H_2|, \lambda_{top}|H_2|,
g_2|L_i|, M_S)$.
For the same reasons as before the optimum choice for the $e_j$ slepton is
the third generation one, i.e. $e_j=$ stau. In fact, this turns out to be
the optimum choice for the UFB-3 direction (note e.g. that the second term
in eq.(\ref{Vufb3}) is now proportional to the slepton masses and thus
smaller) and will represent, as we will
see in sect.6, the {\it strongest} one of {\it all} the UFB and CCB constraints
in the parameter space of the MSSM.

\end{description}

This completes the UFB directions and bounds to take into account in the MSSM.

\section{Improved CCB constraints}

These constraints arise from the existence of charge and color
breaking (CCB) minima in the potential deeper than the
realistic minimum. We have already mentioned the ``traditional" CCB
constraint \cite{Frere} of eq.(\ref{frerebound}).
Other particular CCB constraints have been explored in the
literature \cite{Drees,Gunion,Komatsu,Langacker}.
In this section we will perform a complete analysis of
the CCB minima, obtaining a set of analytic constraints that represent the
necessary and sufficient conditions to avoid the dangerous ones. As we will
see, for certain values of the initial parameters, the CCB constraints
``degenerate" into the previously found UFB constraints since the
minima become unbounded from below directions. In this sense, the
following CCB constraints comprise the UFB bounds of the previous
section, which can be considered as special (but extremely important
as we will see in sect.6) limits of the former.

On the other hand,
we will introduce the one-loop radiative
corrections in a consistent way, a fact that has not been properly
considered up to now.
Actually, as has been explained in the Introduction, the radiative
corrections to the potential can be reasonably approximated by zero
{\em provided that} we are evaluating the tree-level potential at
the appropriate scale.
Therefore, it is still possible
to perform the exploration
of the CCB minima by using the tree-level potential. This simplifies
enormously the analysis, which otherwise would be an impossible task.
At the end of the day, however, it is crucial to substitute the
correct scale (for more details see
subsect.4.5).
This procedure will allow us also to re-evaluate the restrictive power
of the ``traditional" CCB constraints
\footnote{For a recent partial
analysis of this issue using the one-loop potential, see ref.\cite{Bordner}.},
which will be shown in sect.6.

\subsection{General properties}

Let us enumerate a number of general facts which are relevant when
one is looking for CCB constraints in the MSSM.
The proof of the properties 1, 3, 5 below is left for the Appendix, giving
here intuitive arguments of their validity.

\begin{enumerate}

\item[{\bf 1}]
The most dangerous, i.e. the deepest, CCB directions in the
MSSM potential involve only one particular trilinear soft
term of one generation (see eq.(\ref{Vsoft})). This can be either of the
leptonic type (i.e. $A_{e_i}\lambda_{e_i}L_{i} H_1 e_{i}$) or the
hadronic type (i.e. $A_{u_i}\lambda_{u_i}Q_{i} H_2 u_{i}$ or
$A_{d_i}\lambda_{d_i} Q_{i} H_1 d_{i}$). Along one of these particular
directions the remaining trilinear terms are vanishing or negligible.
This is because the presence of a non-vanishing trilinear term in the
potential gives a net negative contribution only in a region of the
field space where the relevant fields are of order $A/\lambda$ with
$\lambda$ and $A$ the corresponding Yukawa coupling
and soft trilinear coefficient;
otherwise either the (positive) mass terms or the (positive)
quartic F--terms associated with these fields dominate the potential.
In consequence two trilinear couplings with different values of
$\lambda$ cannot efficiently ``cooperate" in any region of the field
space to deepen the potential.
Accordingly, to any optimized CCB constraint there corresponds
a unique relevant trilinear coupling.

\item[{\bf 2}]
One cannot say a priori which trilinear coupling gives the
strongest constraints. In particular, contrary to what was claimed in
\cite{Gunion} and used in \cite{Langacker},
it is not true that the trilinear
terms with bigger Yukawa couplings are the most important ones.
This is easy to understand since, although the (negative\footnote{
Recall that the phases of the fields can always be taken so that
the trilinear scalar terms in (\ref{Vsoft}) are negative.}) trilinear
terms, e.g. $A_{u_i}\lambda_{u_i}Q_{i} H_2 u_{i}$, are in principle
more important for larger $\lambda_{u_i}$ couplings, the (positive)
quartic terms, $\lambda_{u_i}^2 \left\{
|Q_{i} u_{i}|^2+|Q_{i} H_2|^2+|H_2 u_{i}|^2\right\}$, are more important
too. So there is a balance and one cannot predict which coupling size,
large or small, will give the most restrictive constraint. We have
examples in both senses.

\item[{\bf 3}]
If the trilinear term under consideration has a Yukawa coupling
$\lambda^2\ll 1$, which occurs in all the cases except for the top, then
along the corresponding deepest CCB direction the D-term must be
vanishing or negligible. Although this may seem quite intuitive,
some authors, particularly in ref.\cite{Gunion}, have argued that
by taking VEVs of the $u_L$ and $u_R$ squarks much smaller than that
of $H_2$, and other fields VEVs being zero (so that the $SU(2)_L\times
U(1)_Y$ D-term is non-vanishing), a non-trivial CCB constraint
appears.
The trouble of
their argument is that they fix $H_1 = 0$
by hand. However, this does not occur neither in the realistic minimum
nor, necessarily, in any optimized CCB direction. We have redone their
analysis in this point, allowing $H_1$ to participate in the
game. Then, one obtains a modified constraint (that substitutes the
one written in eq.(23) of ref.\cite{Gunion}), which turns out to be
equivalent to require positive physical masses for the $u_L$ and
$u_R$ squarks (for more details see the Appendix).

\item[{\bf 4}]
For a given trilinear coupling under consideration
there are {\em two} different relevant directions
to explore. Next, we
illustrate them taking the trilinear coupling of the first
generation,
$A_{u}\lambda_{u}Q_u H_2 u_R$, as a guiding example, specifying how
the directions are generalized to the other couplings.

\vspace{0.2cm}
{\em {\bf Direction a)}}
\subequations{
\be
\label{ccb31}
H_2,Q_u,u_R\neq 0
\ee
\be
\label{ccb32}
|d_{L_j}|^2 =|d_{R_j}|^2
\ee
\be
\label{ccb33}
d_{L_j}d_{R_j} = -\frac{\mu}{\lambda_{d_j}} H_2
\ee
\be
\label{ccb34}
H_1 = 0 \;\;\;{\rm or}\;\;{\rm negligible}
\ee
\be
\label{ccb35}
{\rm Possibly}\;\;\;L_i\neq 0
\ee}
\endsubequations
where $Q_u$ takes the VEV along the $u_L$ direction
and $d_{L_j}$, $d_{R_j}$ are $d$--type squarks such that
\be
\label{lambdas}
\lambda_{d_j}\gg\lambda_u \;\;,
\ee
and whose VEVs are chosen to cancel the $H_1$ F--term
\be
\label{H1Fterm}
\left|\frac{\partial W}{\partial H_1}\right|^2=\left|\mu H_2+
\lambda_{d_j}d_{L_j}d_{R_j}\right|^2= 0\;\;.
\ee
{}From (\ref{lambdas}) and (\ref{ccb33}) it
follows\footnote{The VEVs of the $H_2,Q_u,u_R$ fields are always of
order $A_u/\lambda_u$, as we will see below.}
that $|d_{L_j}|^2\ll |H_2|^2,|Q_u|^2,|u_R|^2$, thus the contribution
of $d_{L_j}, d_{R_j}$ to the D--terms  and the mass soft--terms is negligible.
The net effect of the $d_{L_j},d_{R_j}$ VEVs of eqs.(\ref{ccb32},\ref{ccb33})
is therefore to decrease the $H_2$ squared mass from $m_2^2$
to\footnote{Note that $m_2^2-\mu^2$ is simply the soft mass of $H_2$,
since in the definition of $m_2^2$ is also absorved the $H_1$ F--term,
$|\mu H_2|^2$ (see sect.2).} $m_2^2-\mu^2$.
This interesting fact was first observed in ref.\cite{Komatsu}.
The same job of the $d_j$ squarks can be done by $e_{L_j}$, $e_{R_j}$
sleptons provided that $\lambda_{e_j}\gg\lambda_u$.
$H_1$ must be very small or vanishing, [eq.(\ref{ccb34})],
otherwise the (positive) $d_{L_j}$ and $d_{R_j}$ F--terms,
$\lambda_{d_j}^2 \left\{ |H_1 d_{R_j}|^2+|d_{L_j} H_1|^2\right\}$,
would clearly dominate the potential.
Note that this is also in agreement with the mentioned property 1, i.e.
along a relevant CCB direction in the field-space
only one trilinear scalar coupling can be non-negligible.

In addition to $H_2,Q_u,u_R, d_{L_j},d_{R_j}$, other fields could take
extra non-vanishing VEVs, but as in the above-explained UFB-2 direction
(see sect.3) and for similar reasons, it turns out that the optimum
choice is $L_i\neq 0$, eq.(\ref{ccb35}), with the VEV along the
$\nu_L$ direction (this was not considered in
ref.\cite{Komatsu}).
As we will see, in some special cases $\nu_L\neq 0$ can be
advantageously replaced by\footnote{$e_L$, $e_R$ can be chosen from
different generations in order to avoid the appearance of extra
quartic F-terms. Alternatively, if $\lambda_u\gg \lambda_e$ (as
happens if the lepton is of the first generation and $\tan \beta
\simlt 3$) these new F-terms are negligible. Working under the assumption
of universality of the soft terms both choices are equivalent.}
$e_L\neq 0$, $e_R\neq 0$.
We will not consider this possibility for the moment.

Consequently, the tree-level scalar potential along this {\em (a)}
direction takes the form
\bea
\label{Va}
V &=& \lambda_u^2 \left\{ |H_2 u_R|^2+|Q_u H_2|^2|
+|Q_u u_R|^2\right\}\ +\ {\rm D-terms}
\nonumber \\
&+&
m_{Q_u}^2 |Q_u|^2+m_{u}^2 |u_R|^2
+(m_2^2-\mu^2) |H_2|^2+m_{L_i}^2 |L_i|^2
\nonumber \\
&+&
\left(A_{u}\lambda_{u}Q_u H_2 u_R + {\rm h.c.}\right)\;\; ,
\eea
where we have neglected the contribution of $d_{L_j},d_{R_j}$
to the mass and D terms.

The generalization of this {\em (a)} direction to other couplings
different from\linebreak $A_u\lambda_{u}Q_uH_2u_R$ is as follows.
If the trilinear term under consideration is the charm one, i.e.
$A_{c} \lambda_{c} Q_c H_2 c_R$, everything works as before with
the obvious replacement $u\rightarrow c$
in eqs.(35--\ref{Va}). For the top trilinear term, however,
this direction cannot be applied, since eq.(\ref{lambdas}) cannot
be fulfilled.
If the trilinear term is of the $A_{d_k} \lambda_{d_k} Q_k H_1 d_k$
type, everything is similar interchanging
$H_2$ by $H_1$ and $u$ by $d_k$.
As we will see, for these couplings the presence of an extra VEV
for a slepton occurs normally along the $e_L\neq 0$,
$e_R\neq 0$ direction rather than $\nu_L\neq 0$. In any case, the sleptons
must be chosen from generations satisfying $\lambda_{e_i}\ll\lambda_{d_k}$
in order to make the quartic F-terms associated with them negligible
(this choice is always possible). Let us also note that the above
consideration for the top trilinear coupling is analogously
applicable for the bottom if $\tan\beta\simgt 4$. Finally, the
direction {\em (a)} is generalized to the leptonic couplings,
$A_{e_k} \lambda_{e_k} L_k H_1 e_k$, in a similar way to that of
the $A_{d_k} \lambda_{d_k} Q_k H_1 d_k$ couplings. Now of course
the role of the possible extra leptonic VEVs must be played
by other sleptons, say $L_i'$, $e'_{R_i}$, from a lower generation
than the leptonic coupling under consideration. This excludes
the possibility of extra leptonic VEVs if the latter
corresponds to the electron.

\vspace{0.2cm}
{\em {\bf Direction b)}}
\subequations{
\be
\label{ccb36}
H_2,Q_u,u_R,H_1\neq 0 \;\;,
\ee
\be
\label{ccb37}
{\rm Possibly}\;\;\;L_i\neq 0 \;\;,
\ee}
\endsubequations
where $Q_u$ takes the VEV along the $u_L$ direction.
Note that, according to the general property 1 (see also
Appendix), once we allow $H_1$ to participate in the
game, as reflected in eq.(\ref{ccb36}),
the remaining squark and slepton fields,
apart from those involved in the trilinear coupling, must be vanishing.
The only possible exception is again a slepton $L_i$ VEV along the
$\nu_L\neq 0$ direction. (Or, in some special cases, along the
$e_L, e_R \neq 0$ direction. Then, since $H_1\neq 0$, the associated
Yukawa coupling must satisfy $\lambda_{e_i}\ll\lambda_u$ in order not
to generate extra quartic F-terms; this requires $\tan \beta\simlt 3$.)

Therefore, along the {\em (b)} direction the tree-level scalar potential
takes the form
\bea
\label{Vb}
V &=& \lambda_u^2 \left\{ |H_2u_R|^2 + |Q_uH_2|^2 + |Q_uu_R|^2\right\}\
\nonumber \\
&+& \left(\mu \lambda_{u} Q_u u_R H_1^* + {\rm h.c.}\right)\ +\ {\rm D-terms}
\nonumber \\
&+& m_{Q_u}^2 |Q_u|^2 + m_{u}^2 |u_R|^2 + m_2^2 |H_2|^2
 + m_1^2|H_1|^2 + m_{L_i}^2 |L_i|^2
\nonumber \\
&+& \left( A_{u} \lambda_{u} Q_u H_2 u_R + {\rm h.c.}\right)\
+\ \left(\mu B H_1 H_2+ {\rm h.c.}\right)\;\; .
\eea
Notice that $\left(\mu\lambda_{u}Q_uu_RH_1^*+{\rm h.c.}\right)$
is a piece of the $H_2$ F-term, $|\partial W/\partial H_2|^2$.
Recall that the $|\mu H_1|^2$ piece of this F-term has been absorved in
the definition of $m_1^2$ (see sect.2).

The direction {\em (b)} is generalized to the other trilinear
couplings in a similar way as it was done for direction {\em (a)}.
Let us mention that when dealing with these remaining couplings
there are no restrictions at all on the value of $\tan \beta$.
{}From previous arguments, for the top coupling the direction {\em (b)}
is the only one to be taken into account.


\item[{\bf 5}]
Let us finally comment on the choice of the phases of the various
fields involved in the previous {\em (a)} and {\em (b)} directions.
Again, we continue using the trilinear coupling of the first generation
$A_{u} \lambda_{u} Q_u H_2 u_R$ as a guiding example, but the following
statements are trivially generalized to the other couplings.

If $H_1=0$, i.e. direction {\em (a)}, it is easy to see from (\ref{Va})
that
the only term in the potential without a well-defined phase is the trilinear
scalar term.
Obviously, the fields involved in the coupling can take phases so that
it becomes negative without altering other terms in (\ref{Va}).
This clearly corresponds to the deepest direction in the field-space.
Then, in eq.(\ref{Va}), we can write the trilinear term as
\be
\label{Pha}
-2 |A_{u}\lambda_{u} Q_u H_2 u_R| \;\;\;.
\ee

\vspace{0.2cm}
\noindent
If $H_1\neq 0$ (direction {\em (b)}) there are clearly three terms in the
potential of  eq.(\ref{Vb}) whose phases are in principle undetermined.
These can be written as
\be
\label{Phb}
 2 |A_{u} \lambda_{u} Q_u H_2 u_R| \cos\varphi_1\
+\ 2|\mu \lambda_{u} Q_u H_1 u_R| \cos\varphi_2\
+\ 2|\mu B H_1 H_2 | \cos\varphi_3,
\ee
where $\varphi_i$ are obvious combinations of the signs of
$A_u,B,\mu,\lambda_u$ and the phases of the fields.
Note that $\varphi_1$, $\varphi_2$, $\varphi_3$ are correlationated
parameters.
Now, it can be shown (see Appendix) that
\begin{itemize}

\item
If sign$(A_u)=-$sign$(B)$, the three terms can be made negative
simultaneously, so that after a convenient redefinition of the fields
we can take $\varphi_1=\varphi_2=\varphi_3=\pi$.

\item
If sign$(A_u)=$ sign$(B)$ the previous choice is no longer possible.
Then, for the vaste majority of the cases the deepest direction in
the ($\varphi_1, \varphi_2, \varphi_3$) space corresponds to take
$\varphi_i=\varphi_j=\pi$, $\varphi_l=0$, where $\varphi_l$
corresponds to the smallest term (in absolute value) in eq.(\ref{Phb})
and $\varphi_i,\varphi_j$ are the other two angles.
For the remaining cases this always corresponds to a direction very
close to the deepest one.

\end{itemize}

\end{enumerate}

\subsection{CCB constraints associated with the $Q_u H_2 u_R$ coupling}

Using the previous general properties it is possible to completely
classify the CCB constraints in the MSSM.
According to property 1, there can only be one relevant trilinear
coupling associated to an optimized CCB constraint.
Now, as we did in the previous property 4, we will take the trilinear
coupling of the first generation, $A_{u} \lambda_{u} Q_u H_2 u_R$,
as a guiding example to explain the associated CCB bounds, specifying
how they are generalized to the other couplings.

The bounds arise from the previously expounded {\em (a)} and
{\em (b)}--directions, see eqs.(35) and (39) respectively.
For a given choice of the initial parameters $m,M,A,B,\mu,\lambda_{top}$,
compatible with electroweak breaking and $M_{\rm top}^{\rm exp}$, one can in
principle write down the scalar potential (either eq.(\ref{Va}) or
eq.(\ref{Vb})) at any scale and directly minimize it with respect to
the scalar fields involved.
Then, the possible CCB minima arising should be compared to the realistic
minimum (\ref{Vreal}) in order to decide what is the deepest one.
Of course, all this should be performed at the correct scale in order to
incorporate the radiative corrections properly (recall that this scale
depends itself on what are the relevant VEVs of the fields at the CCB
minimum under consideration).

Unfortunately, despite the form of the potential in eqs.(\ref{Va}),
(\ref{Vb}) is much simpler than the general expression of
eq.(\ref{Vo}), it is still not possible to implement the previous
program in a complete analytical way.
The outcoming equations are in general so involved that they become
useless for practical purposes.
Alternatively, one could follow a numerical procedure,
trying to find out (for each choice of the initial parameters)
the corresponding CCB minima.
This is, however, quite dangerous since there is still a considerable
number of independent variables and the minima usually emerge from subtle
cancellations between different terms, something that can easily escape a
standard program of numerical minimization.
In addition, with the numerical approach the final form for the CCB
bounds is very uneasy to handle and we lose the track of the physical
reasons behind it.
Fortunately, it becomes now feasible to go quite far in the analytic
examination of the general CCB minima, in some cases until the very end
of the analysis, thus obtaining very useful constraints expressed in an
analytical way.
This is the kind of approach we have followed in the paper.
As we will see, the final implementation of these constraints usually
requires a complementary, but trivial, numerical task, namely the scanning
of a certain variable in the range [0,1].

In order to write the CCB constraints it is helpful to express the
various VEVs in terms of the $H_2$ one, using the following notation
\cite{Gunion}
\bea
\label{alfabeta}
|Q_u|&=&\alpha |H_2|\;,\; \; |u_R|=\beta |H_2| \;\;,
\nonumber \\
|H_1|&=&\gamma |H_2|\;,\; \; |L_i|=\gamma_L |H_2| \;\; .
\eea
E.g. the ``traditional" direction, eq.(\ref{frerevevs}), is recovered for the
particular values $\alpha=\beta=1$, $\gamma=\gamma_L=0$.

We shall write now the form of the potential for the directions {\em (a)},
{\em (b)}, obtaining from its minimization the general form of the CCB bounds.
It is convenient for this task to start with the {\em (b)} direction in the
sign$(A_u)=-$sign$(B)$ case,
extending at the end the results to the sign$(A_u)=$ sign$(B)$ case and
to the {\em (a)} direction. The scalar potential
along the direction {\em (b)}, see eq.(\ref{Vb}), can be expressed as
\bea
\label{Valfabeta}
V = \lambda_u^2 F(\alpha,\beta,\gamma,\gamma_L) \alpha^2 \beta^2 |H_2|^4
 -2 \lambda_u \hat A(\gamma) \alpha \beta |H_2|^3
 + {\hat m}^2(\alpha,\beta,\gamma,\gamma_L) |H_2|^2 \;\;,
\eea
where
\bea
\label{FfAm}
F (\alpha,\beta,\gamma,\gamma_L) &=&1+\frac{1}{\alpha^2} +\frac{1}{\beta^2}
+\frac{f(\alpha,\beta,\gamma,\gamma_L)} {\alpha^2\beta^2}\;\;,
\nonumber \\
f(\alpha,\beta,\gamma,\gamma_L)&=& \frac{1}{\lambda_u^2}\left\{
\frac{1}{8} g_2^2 \left(1-\alpha^2-\gamma^2-\gamma_L^2\right)^2\right.
\nonumber \\
\;\;&\ & + \left.
\frac{1}{8}{g'}^2\left(1+\frac{1}{3}\alpha^2-\frac{4}{3}\beta^2
-\gamma^2-\gamma_L^2\right)^2\ +\
\frac{1}{6}g_3^2\left(\alpha^2-\beta^2\right)^2
\right\}\;\;,
\nonumber \\
\hat A(\gamma)&=& |A_u|+|\mu|\gamma\;\;,
\nonumber \\
{\hat m}^2 (\alpha,\beta,\gamma,\gamma_L) &=& m_2^2+m_{Q_u}^2\alpha^2
+m_{u}^2\beta^2 + m_1^2\gamma^2 + m_{L_i}^2\gamma_L^2 - 2|m_3^2|\gamma
\;\; .
\eea
(The $L_i$ VEV has been taken along the direction $\nu_L$ since
otherwise the D--terms cannot be eventually cancelled.)
Then, minimizing $V$ with respect to $|H_2|$ for fixed values of
$\alpha,\beta,\gamma,\gamma_L$, we find, besides the $|H_2|=0$ extremal
(all VEVs vanishing), the following CCB solution
\bea
\label{H2min}
|H_2|_{ext}=|H_2(\alpha,\beta,\gamma,\gamma_L)|_{ext}=
\frac{3\hat A}{4\lambda_u\alpha\beta F}
\left\{1\ +\ \sqrt{1-\frac{8\hat m^2 F}{9\hat A^2}}\;\right\}\;.
\eea
It is easy to check that the solution with a minus sign in front of
the square root in the previous equation corresponds
to a maximum.
Let us note that, as was stated above (see property 1 and
footnote 6), the typical VEVs at a CCB minimum are indeed
of order $A/\lambda$.
The corresponding value of the potential
is
\bea
\label{VCCBmin}
V_{\rm CCB\ min}
=-\frac{1}{2}\alpha\beta |H_2|_{ext}^2\left(\hat A \lambda_u |H_2|_{ext}
-\frac{\hat m^2}{\alpha\beta}\right)\;\;.
\eea
Eqs.(\ref{Valfabeta}--\ref{VCCBmin}) generalize those obtained in
ref.\cite{Gunion}.

Since the  trilinear term of our guiding example has small coupling,
$\lambda_u^2\ll 1$, according to the above property 3 the D--terms
should vanish. This implies
\subequations{
\be
\label{Dcero1}
\alpha^2-\beta^2=0\;,
\ee
\be
\label{Dcero2}
1-\alpha^2-\gamma^2-\gamma_L^2=0 \;\; .
\ee}
\endsubequations
As a consequence $f(\alpha,\beta,\gamma,\gamma_L)$ becomes
vanishing and $F=1+\frac{2}{\alpha^2}$.
Let us note that eq.(\ref{Dcero2}) can only be fulfilled if
$1-\alpha^2-\gamma^2\geq 0$.
In fact, playing only with the $H_2,Q_u,u_R,H_1,L_i$ fields this
is a necessary condition to cancel the D--terms.
If $1-\alpha^2-\gamma^2 < 0$ the cancellation can only be
achieved by including additional fields.
By inspection, the best choice is to take the $L_i$ VEV
along the $e_L$ direction plus an additional
VEV $e_{R_j}=e_{L_i}$.
Then the D--terms are cancelled and
eq.(\ref{Dcero2}) becomes
\be
\label{Dcero3}
1-\alpha^2-\gamma^2+\gamma_L^2=0 \;\; .
\ee
In this case one has to replace $m_{L_i}^2$ by $m_{L_i}^2+m_{e_j}^2$ in
the definition of ${\hat m}^2$, eq.(\ref{FfAm}).
We will not consider this possibility for the moment postponing
for later the discussion of the only situation in which it could be relevant.

The previous CCB minimum, eq.(\ref{VCCBmin}), will be
negative\footnote{The mere existence of a CCB minimum is discarded by
demanding ${\hat A}^2 < (8/9)F\hat m^2$, see eq.(\ref{H2min}).}
unless ${\hat A}^2 \leq F\hat m^2$, i.e.
\bea
\label{GenCond}
\left(|A_u|+|\mu|\gamma\right)^2\ \leq \left(1+\frac{2}{\alpha^2}\right)
\left[ m_2^2+(m_{Q_u}^2+m_{u}^2)\alpha^2
+ m_1^2\gamma^2 + m_{L_i}^2\gamma_L^2 - 2|m_3^2|\gamma
\right]
\eea
where for convenience we have explicitly kept the dependence in
the three variables $\alpha,\gamma,\gamma_L$, which are subject
to eq.(\ref{Dcero2}).
Since $\lambda_u^2\ll 1$, if (\ref{GenCond}) were not satisfied the
corresponding CCB minimum of eq.(\ref{VCCBmin})
would be much deeper ($\propto -1/\lambda_u^2$)
than the realistic one ($\propto -1/g_2^2$), eq.(\ref{Vreal}).
Consequently, eq.(\ref{GenCond}) is the {\em general form}
of the CCB bound
for the {\em (b)}--direction when sign$(A_u)=-$sign$(B)$ and the Yukawa
coupling is much smaller than one, as it is the case at hand.
Let us remark that (\ref{GenCond}) should be satisfied for {\em any}
choice of $\alpha,\gamma,\gamma_L$ obeying eq.(\ref{Dcero2}).
E.g. the ``traditional" bound, eq.(\ref{frerebound}), is recovered for
the particular choice $\alpha=1, \gamma=\gamma_L=0$.

When sign$(A_u)=$ sign$(B)$ one of the three terms
$\left\{|A_u|,|\mu|\gamma,-2|m_3^2|\gamma\right\}$ in eqs.(\ref{FfAm},
\ref{GenCond}) must flip the sign (see property 5 of the previous subsection).

For the {\em (a)}--direction all the equations (\ref{Valfabeta}-\ref{GenCond})
hold making $\gamma=0$, $m_2^2\rightarrow m_2^2-\mu^2$.
In particular eq.(\ref{GenCond}) with these replacements, i.e.
\bea
\label{GenConda}
|A_u|^2\ \leq \left(1+\frac{2}{\alpha^2}\right)
\left[ m_2^2-\mu^2+(m_{Q_u}^2+m_{u}^2)\alpha^2
+ m_{L_i}^2\gamma_L^2 \right] \ ,
\eea
represents the general form of the CCB bounds for direction {\em (a)}.

Clearly, the strongest CCB constraints from (\ref{GenCond}) and
(\ref{GenConda}) arise for
particular values of $\alpha,\gamma,\gamma_L$, which, in turn, depend
on what are the values of various parameters involved in the expressions.
This allows us to be more explicit about the final analytical
form of the CCB constraints and to classify them below:

\begin{description}

\item[CCB-1]
${}^{}$\\
This bound arises by considering the direction {\em (a)} and thus the
general condition (\ref{GenConda}). Then the strongest constraint is obtained
by minimizing the right hand side of (\ref{GenConda}) with respect to
$\alpha$, keeping $\gamma_L^2=1-\alpha^2$.
This gives the following

\begin{enumerate}

\item If $ m_2^2-\mu^2+m_{L_i}^2>0$ and
$3m_{L_i}^2-(m_{Q_u}^2+m_{u}^2)+2(m_2^2-\mu^2) > 0$,
then the optimized CCB-1 bound occurs for $\alpha=1$,
$\gamma_L=0$, i.e.
\bea
\label{CCB1a}
|A_u|^2 \leq 3 \left[ m_2^2-\mu^2+m_{Q_u}^2+m_{u}^2\right]
\eea

\item If $ m_2^2-\mu^2+m_{L_i}^2>0$ and
$3m_{L_i}^2-(m_{Q_u}^2+m_{u}^2)+2(m_2^2-\mu^2) < 0$,
then the optimized bound occurs for
$\alpha, \gamma_L\neq 0$, namely
\bea
\label{CCB1b}
|A_u|^2 \leq \left(1+\frac{2}{\alpha^2}\right)
\left[ m_2^2-\mu^2+(m_{Q_u}^2+m_{u}^2)\alpha^2
+ m_{L_i}^2(1-\alpha^2) \right]
\eea
with $\alpha^2=\sqrt{\frac{2(m_{L_i}^2+m_2^2-\mu^2)}
{m_{Q_u}^2+m_{u}^2-m_{L_i}^2}}$, $\ \gamma_L^2=1-\alpha^2$.

\item If $ m_2^2-\mu^2+m_{L_i}^2<0$, then the CCB-1 bound is automatically
violated since there are many values of $\alpha$ that make the right
hand side of (\ref{GenConda}) negative. In fact the minimization of the
potential in this case gives $\alpha^2 \rightarrow 0$, and we are exactly
led to the UFB-3 direction explained in sect.3, which represents the correct
analysis in this instance.

\end{enumerate}
\noindent
Let us mention that the bound (\ref{CCB1a}) was first obtained in
ref.\cite{Komatsu}. However it seldom represents the optimized
bound, as long as the condition for this (see above eq.(\ref{CCB1a}))
will not normally be satisfied. Hence, eq.(\ref{CCB1b}) will usually
represent the (optimized) CCB-1 bound. Needless to say that the
CCB-1 bound is always stronger than the ``traditional" CCB bounds
\cite{Frere}, see eq.(\ref{frerebound}).

Finally, in the very unlikely case that
$3(m_{L_i}^2+m_{e_j}^2)+(m_{Q_u}^2+m_{u}^2)-2(m_2^2-\mu^2)<0$,
which only can take place in (very strange) non-universal cases, then the
CCB-1 bound would be given by
\bea
\label{CCB1c}
|A_u|^2 \leq \left(1+\frac{2}{\alpha^2}\right)
\left[ m_2^2-\mu^2+(m_{Q_u}^2+m_{u}^2)\alpha^2
+ (m_{L_i}^2+m_{e_j}^2)(\alpha^2-1) \right]
\eea
with $\alpha^2=\sqrt{\frac{2(m_2^2-\mu^2-m_{L_i}^2-m_{e_j}^2)}
{m_{Q_u}^2+m_{u}^2+m_{L_i}^2+m_{e_j}^2}}$, $\gamma_L^2=\alpha^2-1$.

\item[CCB-2]
${}^{}$\\
This bound arises from direction {\em (b)}, i.e. $\gamma\neq 0$, when
sign$(A_u)=-$sign$(B)$. The corresponding CCB constraint
is given by (\ref{GenCond})
with $\gamma_L^2=1-\alpha^2-\gamma^2$, that is
\bea
\label{CCB2}
\hspace{0.3cm}
\left(|A_u|+|\mu|\gamma\right)^2\ \leq
\left(1+\frac{2}{\alpha^2}\right)
\left[ \right. && \hspace{-0.7cm} m_2^2  +
( m_{Q_u}^2+m_{u}^2)\alpha^2 + m_1^2\gamma^2
\nonumber\\
 &&+\; m_{L_i}^2(1-\alpha^2-\gamma^2) - 2|m_3^2|\gamma \left. \right]
\eea
which should be handled in the following way:

\begin{enumerate}

\item Scan $\gamma$ in the range $0\leq\gamma\leq 1$

\item For each value of $\gamma$ the optimum value of $\alpha^2$, i.e.
the one that minimizes the right hand side of (\ref{CCB2}), is in
principle given by
\bea
\label{CCB2p}
\alpha_{ext}^4=\frac{2\left[m_2^2+ m_1^2\gamma^2 + m_{L_i}^2(1-\gamma^2)
- 2|m_3^2|\gamma\right]}{m_{Q_u}^2+m_{u}^2-m_{L_i}^2}
\eea
Under the assumption of universality the denominator of (\ref{CCB2p})
is always positive.
On the other hand, the numerator should also be positive, otherwise the
optimum value of $\alpha$ is $\alpha\rightarrow 0$ and we are exactly
led to the UFB-2 direction explained in sect.3.

\item If $\alpha_{ext}^2<1-\gamma^2$, then $\alpha_{ext}^2$
is indeed the optimum
value of $\alpha^2$ to be substituted in (\ref{CCB2}).

\item  If $\alpha_{ext}^2>1-\gamma^2$, then the D--terms cannot be
cancelled with $\alpha=\alpha_{ext}$ [see eq.(\ref{Dcero2})].
This could be in principle circumvected by including a VEV for
the $e_{R_j}$ slepton, as explained around eq.(\ref{Dcero3}).
Then $\gamma_L^2=\alpha^2+\gamma^2-1$ and the
$m_{L_i}^2(1-\alpha^2-\gamma^2)$ term in (\ref{CCB2}) must be replaced
by $(m_{L_i}^2+m_{e_j}^2)(\alpha^2+\gamma^2-1)$.
The new optimum value of $\alpha_{ext}$ would be in principle given by
\bea
\label{CCB2pp}
\alpha_{ext}'^4=\frac{2\left[m_2^2+ m_1^2\gamma^2 -
(m_{L_i}^2+m_{e_j}^2)(1-\gamma^2)
- 2|m_3|^2\gamma\right]}{m_{Q_u}^2+m_{u}^2+m_{L_i}^2+m_{e_j}^2}
\eea
If $\alpha_{ext}'^2>1-\gamma^2$, then $\alpha_{ext}'^2$ is indeed
the optimum value of $\alpha^2$ to be substituted in (\ref{CCB2})
together with the previous replacements.
If $\alpha_{ext}'^2<1-\gamma^2$, then
the optimum value of $\alpha^2$ is simply $\alpha^2=1-\gamma^2$
(which is equivalent to $\gamma_L=0$), which should be substituted in
(\ref{CCB2}).

\end{enumerate}

\item[CCB-3]
${}^{}$\\
This bound, that also arises from direction {\em (b)}, is to be applied
when\linebreak sign$(A_u)=$ sign$(B)$. It takes exactly the same form
as the CCB-2 one (see above), but flipping the sign of one of the three
terms $\left\{|A_u|,|\mu|\gamma,-2|m_3^2|\gamma\right\}$ in (\ref{CCB2}).
Notice that, due to the form of (\ref{CCB2}) flipping the sign of
$|A_u|$ or the sign of $|\mu|\gamma$ leads to the same result.
Therefore, there are only two choices to examine: the first one writing
$\left(|A_u|-|\mu|\gamma\right)^2$ in the left hand side of (\ref{CCB2}),
the second one writing $+2|m_3^2|\gamma$ in the right hand side of
(\ref{CCB2}) and
hence in those of (\ref{CCB2p}) and (\ref{CCB2pp}).

(Since one cannot know a priori what of the terms listed in
eq.(\ref{Phb}) is going to have the smallest absolute value at the CCB
minimum, one cannot be sure from the beginning which one of the two
choices will be the optimum one.
Consequently, the fastest way to handle this is simply to perform the
examination twice.)

\end{description}

\noindent
Let us finish this subsection by noting that none of the previous CCB
bounds depend on the size of the Yukawa coupling $\lambda_u$ (except
for the fact that $\lambda_u\ll 1$ has been assumed).
However this fact will change as soon as we estimate the appropriate
scale, $Q$, to evaluate them because the size of the tipical VEVs in
the CCB minimum does depend on $\lambda_u$, see eq.(\ref{H2min}).
This issue will be examined in subsect.4.5.

\subsection{Generalization to other couplings}

The previous bounds CCB-1 -- CCB-3 can be straightforwardly generalized
to all the couplings with coupling constant $\lambda\ll 1$.
This includes all the couplings apart from the top.
There are however slight differences depending on the Higgs field
($H_1$ or $H_2$) involved in the coupling.
Thus we expose the various generalizations in a separate way.

\vspace{0.3cm}
\noindent  ${\bf \lambda_c Q_c H_2c_R}$

\vspace{0.2cm}
\noindent
The CCB constraints associated with this coupling have exactly the
same form as those for the $\lambda_u Q_uH_2u_R$ coupling, i.e. the
CCB-1 -- CCB-3 bounds, with the obvious replacement $u \rightarrow c$.

\vspace{0.3cm}
\noindent ${\bf \lambda_d  Q_u H_1d_R}$, ${\bf \lambda_s Q_cH_1s_R}$,
${\bf \lambda_b Q_tH_1b_R}$

\vspace{0.2cm}
\noindent
When dealing with these couplings it is convenient to change the notation
(\ref{alfabeta}), expressing all the VEVs in terms of the $H_1$ one, i.e.
\bea
\label{alfabeta2}
|Q_u|, |Q_c|\ {\rm or}\ |Q_t|&=&\alpha |H_1|\;,\;\; \;
|d_R|,|s_R|\ {\rm or}\ |b_R|=\beta |H_1|\;,
\nonumber \\
|H_2|&=&\gamma |H_1|\;,\; \;\;\;\; \;\;\;\;|L_i|=\gamma_L |H_1|\;,
\;\;
\eea
where $Q_u,Q_c,Q_t$ take the VEVs along the $d_L,s_L,b_L$ directions
respectively.
Then, all the results and equations of subsect.4.2, from
eq.(\ref{Valfabeta}) until the end of the subsection, hold
with the following replacements everywhere
\bea
\label{replace}
H_1 &\leftrightarrow& H_2\;,\hspace{2.3cm}
m_{L_i}^2 \leftrightarrow (m_{L_i}^2+m_{e_j}^2)\;,
\nonumber \\
m_1 &\leftrightarrow& m_2\;,\hspace{2.7cm}
u\rightarrow d, s\ {\rm or}\ b\;.
\eea
Note in particular that if $1-\alpha^2-\gamma^2>0$, the cancellation
of the D--terms requires equal VEVs for $L_i$ (along the $e_L$ direction)
and $e_{R_j}$, while if $1-\alpha^2-\gamma^2<0$ the D--terms can be
cancelled just with $L_i\neq 0$ (along the $\nu_L$ direction).
This works exactly in the opposite way to that of the
$\lambda_u Q_u H_2 u_R$ case.

The modifications in the CCB-1 -- CCB-3 bounds can be
straightforwardly obtained. They remain the same
with the previous eq.(\ref{replace})
substitutions.

\vspace{0.3cm}
\noindent ${\bf \lambda_e L_e H_1 e_R}$,
${\bf\lambda_\mu L_\mu H_1 \mu_R}$, ${\bf\lambda_\tau L_\tau H_1 \tau_R}$

\vspace{0.2cm}
\noindent
The CCB bounds from these couplings have essentially the same form as
the just mentioned $d$-type
ones.
All the results and equations of subsect.4.2, from
eq.(\ref{Valfabeta}) until the end of the subsection, hold
with the following replacements
\bea
\label{noreplace}
H_1 &\leftrightarrow& H_2\;,\hspace{3.3cm}
m_1 \leftrightarrow m_2
\nonumber \\
Q &\rightarrow& L\;,\hspace{3.3cm}
m_{L_i}^2 \rightarrow (m_{L_i'}^2+m_{e_j'}^2)\;,
\nonumber \\
u &\rightarrow& e, \mu\ {\rm or}\ \tau\;
\;,\hspace{2.3cm}
(m_{L_i}^2+m_{e_j}^2) \rightarrow m_{L_i'}^2\;.
\eea
Then $L_e,L_{\mu},L_{\tau}$
take the VEVs along the $e_L,\mu_L,\tau_L$ directions
respectively.
The role of the sleptons $L_i$, $e_{R_j}$ in
the previous subsection is played now by two sleptons
$L'_i$, $e'_{R_j}$ of a different generation than
the trilinear coupling under consideration. In the bounds where
both $L'_i$ (along the direction $e_L'$) and $e'_{R_j}$ take
non-vanishing VEVs, the associated Yukawa coupling, say $\lambda_l'$,
must be much smaller than the Yukawa coupling of the trilinear
coupling under consideration, say $\lambda_l$, in order to avoid
the appearance of large F--terms. Obviously this condition can
always be satisfied except when the coupling under consideration
is of the first generation (i.e. the electron one). Then this
kind of extra VEVs cannot be used, so the optimum value for the
``prime'' sleptons is $e_L'=e_R'=0$, i.e. $\gamma_L=0$.

Under the assumption of universality it is
easy to see that the CCB-1 bound will only take place in the
possibility 1 [see condition above eq.(\ref{CCB1a})],
while the CCB-2, CCB-3 bounds
will always occur in the possibility 4 (note that the
denominator of eq.(\ref{CCB2p}) goes to zero).

\subsection{The case of the top}

Much of
the expounded in subsect.4.2 about the $\lambda_u Q_uH_2u_R$
coupling is still valid for the top one. More precisely, the
eqs.(\ref{alfabeta}--\ref{VCCBmin}) {\em hold} with the replacement
$u\rightarrow t$. However, the top trilinear coupling represents
a special case due to have
the largest Yukawa coupling constant, $\lambda_t$.
This is reflected in the three following differences:

\begin{itemize}

\item The D-terms along an optimized CCB direction are no longer
vanishing or negligible, since $\lambda_t=O(1)$, which implies
that the D--terms and the F--terms have orders of magnitude
comparable [see property 3 in sect.4.1]. Consequently, eqs.(48)
or (\ref{Dcero3}) should not be imposed now.

\item The direction {\em (a)} specified in eqs.(35) is no longer
applicable due to the absence of $d$--type squarks such that
$\lambda_{d_j}\gg\lambda_t$. Consequently, the only direction
to take into account is the {\em (b)} one, eqs.(39), and
the CCB-1 bound does not apply to the top case.

\item Since  $\lambda_t=O(1)$ it is no longer true that a negative
minimum ($\propto -1/\lambda_t^2$)
associated to the top trilinear coupling is necessarily
much deeper
than the realistic minimum ($\propto -1/g_2^2$), thus destabilizing the
standard vacuum, as can be easily seen
by examining eqs.(\ref{H2min},\ref{VCCBmin},\ref{Vreal}). Therefore, rather
than the absence of a negative minimum, we must demand that the
possible CCB minimum satisfies $V_{\rm CCB\;min}>V_{\rm real\;min}$,
where  $V_{\rm CCB\;min},V_{\rm real\;min}$ are given by
eqs.(\ref{VCCBmin}),(\ref{Vreal}).

\end{itemize}

In the following we will still use the $SU(3)$ D--term cancellation
condition
\be
\label{Dcero4}
|Q_t|^2=|t_R|^2\;\rightarrow\; \alpha^2=\beta^2\;,
\ee
taking the VEV of $Q_t$ along $t_L$.
This particular direction proves to be very close to the deepest one,
simplifying substantially the subsequent analysis. The analogous
approximation for the $SU(2)\times U(1)_Y$ D--terms is, however, not
good (this comes from the smaller size of the associated gauge couplings), so
we will allow them to be non-vanishing.

Since we have to analyze the potential along the direction {\em (b)},
we must keep in mind that there are two different scenarios
depending on the relative sign of $A_t$ and $B$, see property 5 in
subsect.4.1. In the following we will assume sign$(A_t)=-$sign$(B)$,
which represents the simplest case. The extension of the results
to the sign$(A_t)=$ sign$(B)$ case is trivial and will be given at the end.

{}From (\ref{Valfabeta}, \ref{FfAm}) we can optimize the value of
$\gamma_L=|L_i|/|H_2|$. This is given by
\bea
\label{Lext}
(\gamma_L^2)_{ext}=1-\gamma^2-\alpha^2
-\frac{4 m_{L_i}^2}{(g'^2+g_2^2)|H_2|^2}\;.
\eea
Notice that this value is only acceptable if $(\gamma_L^2)_{ext}>0$,
which, as we shall see, will have to be checked at the end of the
examination. Assuming for the time being that indeed
$(\gamma_L^2)_{ext}>0$, the potential (with
$\gamma_L=(\gamma_L)_{ext}$) is given from eq.(\ref{Valfabeta}) by
\bea
\label{Valfabeta2}
V = \lambda_t^2 F'(\alpha) \alpha^4 |H_2|^4
 -2 \lambda_t \hat A'(\gamma) \alpha^2  |H_2|^3
 + {{\hat m}'^2}(\alpha,\gamma) |H_2|^2
- \frac{2 m_{L_i}^4}{g'^2+g_2^2}\;\;,
\eea
with
\bea
\label{FfAm2}
F'(\alpha) &=&1+\frac{2}{\alpha^2} +\frac{f'}{\alpha^4}\;;\;\;f'=0\;,
\nonumber \\
\hat A'(\gamma)&=& |A_t|+|\mu|\gamma\;\;,
\nonumber \\
{{\hat m}'^2} (\alpha,\gamma) &=&
m_2^2+(m_{Q_t}^2+m_{t}^2)\alpha^2+m_1^2\gamma^2
+m_{L_i}^2(1-\alpha^2-\gamma^2)-2|m_3^2|\gamma.
\eea

This can be handled in the following way:

\begin{enumerate}

\item Scan $\gamma$ in the range $0\leq\gamma\leq 1$

\item For each value of $\gamma$ the optimum values of $\alpha^2$,
$H_2$
 i.e.
the ones that minimize the right hand side of (\ref{Valfabeta2}), are
given by
\bea
\label{alfext}
\alpha_{ext}^2=\frac{\hat A'(\gamma)}{\lambda_t |H_2|_{ext}}
-1 -\frac{m_{Q_t}^2+m_{t}^2-m_{L_i}^2}{2\lambda_t^2 |H_2|_{ext}^2}
\;\;,
\eea
\bea
\label{H2ext}
|H_2|_{ext}=
\frac{3 \hat A'(\gamma)}{4\lambda_t\alpha_{ext}^2 F'(\alpha_{ext})}
\left\{1\ +\ \sqrt{1-\frac{8{{\hat m}'^2}(\alpha_{ext},\gamma)
F'(\alpha_{ext})}{9 \hat A'^2(\gamma)}}\;
\right\}\;\;.
\eea
For each value of $\gamma$ the coupled equations (\ref{alfext}),
(\ref{H2ext}) can be solved, e.g. by a numerical method. Then,
the consistency of the procedure requires
\bea
\label{consist}
\alpha_{ext}^2>0,\;\;\;|H_2|_{ext}>0,\;\;\;(\gamma_L^2)_{ext}>0\;\;,
\eea
where $\alpha_{ext},|H_2|_{ext},(\gamma_L)_{ext}$ are given by
eqs.(\ref{alfext}), (\ref{H2ext}) and (\ref{Lext}) respectively.

If (\ref{consist}) is fulfilled, then the corresponding value of the
potential at the minimum is given by
\bea
\label{VCCBmin2}
V_{\rm CCB\ min}=-\frac{1}{2}\alpha_{ext}^2 |H_2|_{ext}^2
\left(\lambda_t \hat A'(\gamma) |H_2|_{ext}
-\frac{\hat m'^2(\alpha_{ext},\gamma)}{\alpha_{ext}^2}\right)
- \frac{2 m_{L_i}^4}{g'^2+g_2^2}\;\;.
\eea
This value will be negative unless $\hat A'^2 \leq F'\hat m'^2$, i.e.
\bea
\label{NGenCond}
\hspace{0.3cm}
\left(|A_t|+|\mu|\gamma\right)^2\ \leq
\left(1+\frac{2}{\alpha^2}\right)
\left[ \right. && \hspace{-0.7cm} m_2^2  +
( m_{Q_t}^2+m_{t}^2)\alpha^2 + m_1^2\gamma^2
\nonumber\\
 &&+\; m_{L_i}^2(1-\alpha^2-\gamma^2) - 2|m_3^2|\gamma \left. \right] \ .
\eea
E.g. the ``traditional" CCB bound of the type of
eq.(\ref{frerebound}) is recovered for
the particular choice $\alpha=1, \gamma=0$. However, as mentioned above,
a negative minimum associated to the top trilinear coupling is not
necessarily deeper than the realistic minimum.
Consequently, the CCB bound to be imposed has the form
\be
\label{CCBtop}
V_{\rm CCB\ min}> V_{\rm real\;min}
\;\;,
\ee
where $V_{\rm CCB\ min}$ and $V_{\rm real\;min}$ are given by
eqs.(\ref{VCCBmin2}) and (\ref{Vreal}) respectively.

\item If (\ref{consist}) is not fulfilled, this means that there is
no CCB minimum with $\gamma_L=(\gamma_L)_{ext}$. Then, necessarily, the
optimum value of $\gamma_L$ is
\be
\label{L0}
\gamma_L=0
\;\;,
\ee
which implies
\bea
\label{Valfabeta2b}
V = \lambda_t^2 F'(\alpha,\gamma) \alpha^4 |H_2|^4
 -2 \lambda_t \hat A'(\gamma) \alpha^2  |H_2|^3
 + {{\hat m}'^2}(\alpha,\gamma) |H_2|^2 \;\;,
\eea
The optimum values of $\alpha$, $H_2$ are now given by
\bea
\label{alfext2}
\alpha_{ext}^2=\frac{8\lambda_t^2}{g'^2+g_2^2+8\lambda_t^2}
\left[\frac{\hat A'(\gamma)}{\lambda_t|H_2|_{ext}}
-1-\frac{m_{Q_t}^2+m_{t}^2}{2 \lambda_t^2 |H_2|_{ext}^2}
+\frac{g'^2+g_2^2}{8 \lambda_t^2}(1-\gamma^2)\right]
\eea
\bea
\label{H2ext2}
|H_2|_{ext}=
\frac{3 \hat A'(\gamma)}{4\lambda_t\alpha_{ext}^2 F'(\alpha_{ext},\gamma)}
\left\{1\ +\ \sqrt{1-
\frac{ 8 {\hat m}'^2 (\alpha_{ext},\gamma) F'(\alpha_{ext},\gamma) }
{ 9 \hat A'^2(\gamma) }     }\;\right\}\;\;.
\eea
with
\bea
\label{FfAm3}
F'(\alpha,\gamma) &=&1+\frac{2}{\alpha^2} +\frac{f'}{\alpha^4}\;\;,
\nonumber \\
f'&=& \frac{g'^2+g_2^2}{8\lambda_t^2}
\left(1-\alpha^2-\gamma^2\right)^2\;\;,
\nonumber \\
{\hat m}'^2 (\alpha,\gamma) &=& m_2^2+\left(m_{Q_t}^2
+m_{t}^2\right)\alpha^2 + m_1^2\gamma^2 - 2|m_3^2|\gamma
\;\; .
\eea
Consistency now requires
\bea
\label{consist2}
\alpha_{ext}^2>0,\;\;\;|H_2|_{ext}>0\;\;.
\eea
Otherwise there is {\em no} CCB minimum for the particular value
of $\gamma$
being scanned. If (\ref{consist2}) is satisfied, then the value
of the potential at the minimum is given by
\bea
\label{VCCBmin3}
V_{\rm CCB\ min} = -\frac{1}{2} \alpha_{ext}^2 |H_2|_{ext}^2
\left( \hat A'(\gamma) \lambda_t |H_2|_{ext}
- \frac{\hat m'^2(\alpha_{ext},\gamma)}{\alpha_{ext}^2} \right)\;\;.
\eea
and the CCB bound takes again the form
\be
\label{CCBtop2}
V_{\rm CCB\ min}> V_{\rm real\;min}
\;\;.
\ee

\end{enumerate}

When sign$(A_t)=$ sign$(B)$ the analysis is exactly the same but, as
usual,  one of the three terms proportional to
$|A_t|,\ |\mu|\gamma,\ |m_3^2|\gamma$ in eqs.(\ref{FfAm2}),
(\ref{FfAm3}) must flip its sign.

Let us finally note that if $\tan\beta$ is large ($\tan\beta\simgt
15$), then $\lambda_b=O(1)$ and the analysis of this subsection
is also the correct one for the bottom, performing the substitutions
\bea
\label{treplace2}
H_1 &\leftrightarrow& H_2\;, \hspace{2.35cm}m_{L_i}^2\rightarrow
(m_{L_i}^2+m_{e_j}^2)\;,
\nonumber \\
m_1 &\leftrightarrow& m_2\;,\hspace{3cm}t\rightarrow b\;.
\eea

\subsection{The choice of the scale}

As is well known (see e.g. ref.~\cite{ford}) the {\em complete}
effective potential, $V(Q,\lambda_\alpha(Q),\ m_\beta(Q),$
$\phi(Q))$ (in short $V(Q,\phi)$), where $Q$ is the
renormalization scale,
$\lambda_\alpha(Q),m_\beta(Q)$ are running parameters and masses,
and $\phi(Q)$ are the generic classical fields, is scale-independent,
i.e.
\be
\label{Qind}
\frac{d V}{dQ}=0\;\;.
\ee
This property allows in principle a different scale for each value
of the classical fields, i.e. $Q=f(\phi)$. Denoting by
$\langle \phi\rangle$ the VEVs of the $\phi$--fields obtained from
the minimization condition on $V$, it is clear that the two
following minimization conditions
\be
\label{mincond1}
\frac{\partial V(Q=f(\phi),\phi)}{\partial \phi} = 0
\ee
\be
\label{mincond2}
\left.\frac{\partial V(Q,\phi)}{\partial \phi}\right|_{Q=f(\phi)} = 0
\ee
yield equivalent results for $\langle \phi\rangle$ (for a more detailed
discussion see ref.\cite{CEQR}).

\vspace{0.3cm}
\noindent
The previous results apply exactly {\em only} to the complete effective
potential. In practice, however, we can only know $V$ with a
certain degree of accuracy in a perturbative expansion. In particular,
at one-loop level
\be
\label{V1def}
V_1=V_o(Q,\phi) + \Delta V_1(Q,\phi)
\ee
where $V_o$ is the (one-loop improved) tree-level potential
and $\Delta V_1$ is the one-loop radiative correction to the
effective potential
\be
\label{DeltaV1p}
\Delta V_1={\displaystyle\sum_{\alpha}}{\displaystyle\frac{n_\alpha}{64\pi^2}}
M_\alpha^4\left[\log{\displaystyle\frac{M_\alpha^2}{Q^2}}
-\frac{3}{2}\right]\;\;,
\ee
with
$M_\alpha^2(Q)$ being all the
(in general field--dependent) tree-level squared mass eigenstates
(see also eq.(\ref{V1})). $V_1(Q,\phi)$ does {\em not} obey
eq.(\ref{Qind}) for any $Q$, but it is clear that
in the region of $Q$ of the order
of the most significant masses appearing in (\ref{DeltaV1p}), the logarithms
involved in the radiative corrections, and hence the
radiative corrections themselves, are minimized, thus improving the
perturbative expansion. As a matter of fact, in that region of $Q$,
$V_1$ is approximately scale-independent \cite{Gamberini,CC},
so eq.(\ref{Qind}) is nearly satisfied. Consequently, by choosing
an appropriate value of $Q$, eqs.(\ref{mincond1}) and
eq.(\ref{mincond2}), plugging $V\rightarrow V_1$, produce essentially
the same values of $\langle\phi \rangle$, although, of course,
eq.(\ref{mincond2}) is much easier to handle. This statement can
be numerically confirmed, see e.g. ref.\cite{CEQR}.

\vspace{0.3cm}
\noindent
Finally, choosing a $Q$ scale, say $\hat Q$, such that
${\partial\Delta V_1}/{\partial\phi} = 0$, we will get
the same results from eq.(\ref{mincond2}) using
$V_1$ or\footnote{Actually,
this has been our procedure in
sect.2 when analyzing the realistic minimum, $V_{\rm real\;min}$.
We concluded there that a good choice of the scale in order to
evaluate $V_{\rm real\;min}$ was $\hat Q=M_S$, where $M_S$
(a certain average of the relevant $M_\alpha$ masses) was
given by eq.(\ref{MS}).} $V_o$.
On the other hand $\hat Q$ always belongs to the
above-mentioned stability region since at $\hat Q$
the logarithms involved in $\Delta V_1$, and $\Delta V_1$ itself,
are necessarily small, thus optimizing the perturbative expansion.
For the CCB directions the equation
${\partial\Delta V_1}/{\partial\phi} = 0$ amounts to a extremely
involved condition but from the previous arguments it is sufficiently
good for our calculation
to take $\hat Q$ of the order of the most significant
$M_\alpha$ mass appearing in (\ref{DeltaV1p}) (the precise value
is irrelevant), thus suppressing the relevant logarithms,
and then use eq.(\ref{mincond2}) plugging
$V\rightarrow V_o(\hat Q)$. This was also the procedure proposed
in ref.\cite{Gamberini}.

\vspace{0.3cm}
\noindent
Turning back to our specific task, we have to choose the appropriate
scale $\hat Q$ to evaluate the existence of CCB minima in the
potential and the subsequent CCB bounds. Now in eq.(\ref{DeltaV1p}),
besides masses of order $M_S$, there appear other (field-dependent)
masses. In general the latter will be much larger than $M_S$ since
the typical magnitude of the relevant fields in a CCB minimum is
$O(M_S/\lambda)$. A more precise measure of the size of the most
significant masses appearing in (\ref{DeltaV1p}) comes from the
explicit tree-level expresions for the VEVs of the relevant fields
at the CCB minimum (see in particular eq.(\ref{H2min})) and from
the inspection of what $M_\alpha$ masses they give rise to in the $V_o$
potential. In this way we obtain the following estimations of the
size of the appropriate scale, $\hat Q$, depending on the relevant
trilinear coupling associated with the CCB bound under consideration
\bea
\label{escalas}
{\lambda_u  Q_u H_2u_R}, {\lambda_c Q_cH_2c_R},
{\lambda_t Q_tH_2t_R}:\hspace{0.5cm}
\hat Q_{u,c,t}&\sim&{\rm Max}\left(M_S,g_3\frac{A_{u,c,t}}{4\lambda_{u,c,t}},
\lambda_t\frac{A_{u,c,t}}{4\lambda_{u,c,t}}
\right)
\nonumber \\
{\lambda_d  Q_u H_1d_R}, {\lambda_s Q_cH_1s_R},
{\lambda_b Q_tH_1b_R}:\hspace{0.5cm}
\hat Q_{d,s}&\sim&{\rm Max}\left(M_S,g_3\frac{A_{d,s}}{4\lambda_{d,s}}
\right),
\nonumber \\
\hat Q_b&\sim&{\rm Max}\left(M_S,g_3\frac{A_{b}}{4\lambda_{b}},
\lambda_t\frac{A_{b}}{4\lambda_{b}}
\right)
\nonumber \\
{\lambda_e  L_e H_1e_R}, {\lambda_\mu L_\mu H_1\mu_R},
{\lambda_\tau L_\tau H_1\tau_R}:\hspace{0.5cm}
\hat Q_{e,\mu,\tau}&\sim&{\rm Max}\left(M_S,g_2\frac{A_{e,\mu,\tau}}
{4\lambda_{e,\mu,\tau}}\right)
\eea
Moreover, for $\hat Q_{d,s}$, $\hat Q_{e,\mu,\tau}$, if we are considering
the CCB-2,3 bounds, which involve $H_2\neq 0$, we have to include
$\lambda_t\frac{A_{d,s}}{4\lambda_{d,s}}$, $\lambda_t
\frac{A_{e,\mu,\tau}}{4\lambda_{e,\mu,\tau}}$, respectively
among the various quantities within the parenthesis above.

Finally, let us note that a similar procedure for the choice of the
$\hat Q$ scale was carried out in sect.3 for the UFB bounds.

\vspace{0.3cm}
\noindent
Of course, the results for CCB and UFB bounds are quite stable against
moderate variations of the $\hat Q$--scale.

\section{Summary of UFB and CCB constraints}

Here we summarize the two types of constraints, UFB and CCB,
analyzed in sect.3 and sect.4 respectively,
to which the reader is referred for
further details.


\subsection {UFB constraints}

\noindent
These constraints arise from directions in the field-space along which
the (tree-level) potential becomes unbounded from below (UFB). It is
interesting
to note that usually this is only true at tree-level since radiative
corrections eventually raise the potential for large enough values of
the fields. This is the case of UFB-2,3 below.

%
\begin{description}

\item[UFB-1]
${}^{}$\\
The condition
\be
\label{SU1}
m_1^2+m_2^2 \ge 2 |m_3^2|
\ee
must be verified at any scale $Q>M_S$ and, in particular, at the
unification scale
$Q=M_X$. $M_S$ is the typical scale of SUSY masses
(see e.g. eq.(\ref{MS})).

\item[UFB-2]
${}^{}$\\
For any value of $|H_2|<M_X$ satisfying
\be
\label{SU2}
|H_2|^2 > \frac{ 4m_{L_i}^2 }
{ (g'^2+g_2^2) \Big[1-\frac{|m_3|^4}{(m_1^2-m_{L_i}^2)^2} \Big] } \,
\ee
and provided that
\be
\label{SU3}
|m_3^2| < m_1^2-m_{L_i}^2
\ee
the following condition must be verified:
\be
\label{SU4}
V_{\rm UFB-2}(Q=\hat Q) > V_{\rm real \; min}(Q=M_S) \;,
\ee
where $V_{\rm real \; min}$ is the value of the realistic minimum, given
by eq.(\ref{Vreal}),
$\hat Q\sim {\rm Max}(g_2 |H_2|,\ \lambda_{top}|H_2|,\ M_S)$, and
\be
\label{SU5}
V_{\rm UFB-2} = \left[ m_2^2 + m_{L_i}^2
- \frac{|m_3|^4}{m_1^2-m_{L_i}^2}\right] |H_2|^2
- \frac{2m_{L_i}^4}{g'^2+g_2^2}  \ .
\ee

\item[UFB-3]
${}^{}$\\
For any value of $|H_2|<M_X$ satisfying
\be
\label{SU6}
|H_2| > \sqrt{ \frac{\mu^2}{4\lambda_{e_j}^2}
+ \frac{4m_{L_i}^2}{g'^2+g_2^2}}-\frac{|\mu|}{2\lambda_{e_j}} \ ,
\ee
with $j\neq i$ the following condition must be verified:
\be
\label{SU7}
V_{\rm UFB-3}(Q=\hat Q) > V_{\rm real \; min}(Q=M_S) \ ,
\ee
where $V_{\rm real \; min}$ is given by eq.(\ref{Vreal}),
$\hat Q\sim {\rm Max}(g_2 |e|, \lambda_{top} |H_2|,
g_2 |H_2|, g_2 |L_i|, M_S)$
with
$|e|$=$\sqrt{\frac{|\mu|}{\lambda_{e_j}}|H_2|}$ and
$|L_i|^2$=$-\frac{4m_{L_i}^2}{g'^2+g_2^2}$+($|H_2|^2$+$|e|^2$),
$\lambda_{e_j}$ is an $e$-type Yukawa coupling and
\be
\label{SU8}
V_{\rm UFB-3}=(m_2^2 -\mu^2+ m_{L_i}^2 )|H_2|^2
+ \frac{|\mu|}{\lambda_{e_j}} ( m_{L_j}^2+m_{e_j}^2+m_{L_i}^2 ) |H_2|
-\frac{2m_{L_i}^4}{g'^2+g_2^2} \ .
\ee
If $|H_2|$ does not satisfy eq.(\ref{SU6}), the constraint is still
given in the form (\ref{SU7}), but with
\be
\label{SU9}
V_{\rm UFB-3}= (m_2^2 -\mu^2 ) |H_2|^2
+ \frac{|\mu|} {\lambda_{e_j}} ( m_{L_j}^2+m_{e_j}^2 ) |H_2| + \frac{1}{8}
(g'^2+g_2^2)\left[ |H_2|^2+\frac{|\mu|}{\lambda_{e_j}}|H_2|\right]^2 \ .
\ee
{}From (\ref{SU7}), (\ref{SU8}), (\ref{SU9}), it is clear that the larger
$\lambda_{e_j}$ the more restrictive
the constraint becomes. Consequently, the optimum choice of
the $e$--type slepton should be the third generation one, i.e.
${e_j}=$ stau.

It is interesting to mention that the previous constraint (\ref{SU7})
with the following
replacements
\bea
\label{otramas}
e \rightarrow d\;,\hspace{2.0cm}
\lambda_{e_j} \rightarrow \lambda_{d_j}\;,\hspace{2.0cm}
L_j \rightarrow Q_j\;,
\eea
must also be imposed. Now $i$ may be equal to $j$ (the optimum choice is
${d_j}=$ sbottom)
and $\hat Q\sim {\rm Max}\ (g_2 |H_2|,\  \lambda_{top} |H_2|,
\ g_3 |d|,\ \lambda_{u_j} |d|, \ g_2 |L_i|,\ M_S)$.
However, the optimum condition is the first one
with the sleptons (note e.g. that the second term
in eq.(\ref{SU8}) is proportional to the slepton masses and thus
smaller) and will represent, as we will
see in sect.6, the {\it strongest} one of {\it all} the UFB and CCB constraints
in the parameter space of the MSSM.

\end{description}

\subsection{ CCB constraints}

\noindent
These constraints arise from the existence of charge and color breaking (CCB)
minima in the potential deeper than the realistic minimum. As
was explained in subsect.4.1 and Appendix, the most dangerous, i.e. the
deepest, CCB directions in the MSSM potential involve only one particular
trilinear soft term of one generation. Then, for each trilinear soft term
we will write below the three possible (optimized) types of constraints that
emerge. Following the
notation of the previous section, they are named CCB-1,2,3.

\vspace{0.3cm}
\noindent  ${\bf \lambda_u Q_u H_2u_R}$

\vspace{0.3cm}
\noindent
The following constraints must be evaluated at the scale
$\hat Q\sim {\rm Max}\left(M_S, g_3\frac{A_u}{4\lambda_u},
\lambda_t\frac{A_u}{4\lambda_u}\right)$.

\begin{description}

\item[CCB-1]
${}^{}$
\begin{enumerate}
\item If $ m_2^2-\mu^2+m_{L_i}^2>0$ and
$3m_{L_i}^2-(m_{Q_u}^2+m_{u}^2)+2(m_2^2-\mu^2) > 0$,
then the optimized CCB-1 bound is
\bea
\label{CCCB1a}
|A_u|^2 \leq  3 \left[ m_2^2-\mu^2+m_{Q_u}^2+m_{u}^2\right]
\eea

\item If $ m_2^2-\mu^2+m_{L_i}^2>0$ and
$3m_{L_i}^2-(m_{Q_u}^2+m_{u}^2)+2(m_2^2-\mu^2) < 0$,
then the optimized CCB-1 bound is
\bea
\label{CCCB1b}
|A_u|^2 \leq \left(1+\frac{2}{\alpha^2}\right)
\left[ m_2^2-\mu^2+(m_{Q_u}^2+m_{u}^2)\alpha^2
+ m_{L_i}^2(1-\alpha^2) \right]
\eea
with $\alpha^2=\sqrt{\frac{2(m_{L_i}^2+m_2^2-\mu^2)}
{m_{Q_u}^2+m_{u}^2-m_{L_i}^2}}.$

\item If $ m_2^2-\mu^2+m_{L_i}^2<0$, then the CCB-1 bound is automatically
violated. In fact the minimization of the
potential in this case gives $\alpha^2 \rightarrow 0$, and we are exactly
led to the UFB-3 direction shown above, which represents the correct
analysis in this instance.

\end{enumerate}
\noindent
Let us mention that the bound (\ref{CCCB1a}) seldom represents the optimized
bound, as long as the condition for this (see above eq.(\ref{CCCB1a}))
will not normally be satisfied. Hence, eq.(\ref{CCCB1b}) will usually
represent the (optimized) CCB-1 bound.

Finally, in the very unlikely case that
$3(m_{L_i}^2+m_{e_j}^2)+(m_{Q_u}^2+m_{u}^2)-2(m_2^2-\mu^2)<0$,
which only can take place in (very strange) non-universal cases, then the
CCB-1 bound would be given by
\bea
\label{CCCB1c}
|A_u|^2 \leq \left(1+\frac{2}{\alpha^2}\right)
\left[ m_2^2-\mu^2+(m_{Q_u}^2+m_{u}^2)\alpha^2
+ (m_{L_i}^2+m_{e_j}^2)(\alpha^2-1) \right]
\eea
with $\alpha^2=\sqrt{\frac{2(m_2^2-\mu^2-m_{L_i}^2-m_{e_j}^2)}
{m_{Q_u}^2+m_{u}^2+m_{L_i}^2+m_{e_j}^2}}$.

\item[CCB-2]
${}^{}$\\
This second constraint applies whenever sign$(A_u)=-$sign$(B)$.
The general form of the CCB-2 constraint is
\bea
\label{CCCB2}
\hspace{0.3cm}
\left(|A_u|+|\mu|\gamma\right)^2\ \leq
\left(1+\frac{2}{\alpha^2}\right)
\left[ \right. && \hspace{-0.7cm} m_2^2  +
( m_{Q_u}^2+m_{u}^2)\alpha^2 + m_1^2\gamma^2
\nonumber\\
 &&+\; m_{L_i}^2(1-\alpha^2-\gamma^2) - 2|m_3^2|\gamma \left. \right]
\eea
which should be handled in the following way:

\begin{enumerate}

\item Scan $\gamma$ in the range $0\leq\gamma\leq 1$

\item For each value of $\gamma$ the optimum value of $\alpha^2$, i.e.
the one that minimizes the right hand side of (\ref{CCCB2}), is in
principle given by
\bea
\label{CCCB2p}
\alpha_{ext}^4=\frac{2\left[m_2^2+ m_1^2\gamma^2 + m_{L_i}^2(1-\gamma^2)
- 2|m_3^2|\gamma\right]}{m_{Q_u}^2+m_{u}^2-m_{L_i}^2}
\eea
Under the assumption of universality the denominator of (\ref{CCCB2p})
is always positive.
On the other hand, the numerator should also be positive, otherwise the
optimum value of $\alpha$ is $\alpha\rightarrow 0$ and we are exactly
led to the UFB-2 direction explained above.

\item If $\alpha_{ext}^2<1-\gamma^2$, then $\alpha_{ext}^2$ is the optimum
value of $\alpha^2$ to be substituted in (\ref{CCCB2}).

\item  If $\alpha_{ext}^2>1-\gamma^2$ and $\tan\beta\simlt 3$, then the
$m_{L_i}^2(1-\alpha^2-\gamma^2)$ term in (\ref{CCCB2}) must be replaced
by $(m_{L_i}^2+m_{e_j}^2)(\alpha^2+\gamma^2-1)$.
The new optimum value of $\alpha_{ext}$ would be in principle given by
\bea
\label{CCCB2pp}
\alpha_{ext}'^4=\frac{2\left[m_2^2+ m_1^2\gamma^2 -
(m_{L_i}^2+m_{e_j}^2)(1-\gamma^2)
- 2|m_3|^2\gamma\right]}{m_{Q_u}^2+m_{u}^2+m_{L_i}^2+m_{e_j}^2}
\eea
If $\alpha_{ext}'^2>1-\gamma^2$, then $\alpha_{ext}'^2$ is
the optimum value of $\alpha^2$ to be substituted in (\ref{CCCB2})
together with the previous replacement.
If $\alpha_{ext}'^2<1-\gamma^2$, then
the optimum value of $\alpha^2$ is simply $\alpha^2=1-\gamma^2$,
which should be substituted in (\ref{CCCB2}).

\item  If $\alpha_{ext}^2>1-\gamma^2$ and $\tan\beta> 3$, then
the optimum value of $\alpha^2$ is simply $\alpha^2=1-\gamma^2$,
which should be substituted in (\ref{CCCB2}).

\end{enumerate}

\item[CCB-3]
${}^{}$\\
This bound is the equivalent to the CCB-2 one, but when
sign$(A_u)=$ sign$(B)$. It has
exactly the same form as CCB-2 but flipping the sign of one of the three
terms $\left\{|A_u|,|\mu|\gamma,-2|m_3^2|\gamma\right\}$ in (\ref{CCCB2}).
Notice that, due to the form of (\ref{CCCB2}) flipping the sign of
$|A_u|$ or the sign of $|\mu|\gamma$ leads to the same result.
Therefore, there are only two choices to examine: the first one writing
$\left(|A_u|-|\mu|\gamma\right)^2$ in the left hand side of (\ref{CCCB2}),
the second one writing $+2|m_3^2|\gamma$ in the right hand side of
(\ref{CCCB2}) and hence in those of (\ref{CCCB2p}) and (\ref{CCCB2pp}).

\end{description}

\vspace{0.3cm}
\noindent  ${\bf \lambda_c Q_c H_2c_R}$

\vspace{0.3cm}
\noindent
The CCB constraints associated with this coupling have exactly the
same form as those for the $\lambda_u Q_uH_2u_R$ coupling, i.e. the
CCB-1 -- CCB-3 bounds, with the
obvious replacement $u \rightarrow c$ (this is also valid for the scale
$\hat Q$).
Now, there is no constraint on $\tan \beta$
and, therefore, possibility 4 in CCB-2,3
can be applied for
any value of $\tan \beta$ and possibility 5 should not be taken
into account.

\vspace{0.3cm}
\noindent ${\bf \lambda_d  Q_u H_1d_R}$, ${\bf \lambda_s Q_cH_1s_R}$,
${\bf \lambda_b Q_tH_1b_R}$

\vspace{0.3cm}
\noindent
Now the scale is given by:
$\hat Q_{d,s}\sim {\rm Max}\left(M_S, g_3\frac{A_{d,s}}{4\lambda_{d,s}}\right),
\hat Q_b\sim {\rm Max}\left(M_S,g_3\frac{A_{b}}{4\lambda_{b}},
\lambda_t\frac{A_{b}}{4\lambda_{b}}\right)$.

The CCB-1 bounds, eqs.(\ref{CCCB1a},\ref{CCCB1b},\ref{CCCB1c}),
remain the same with the following replacements
\bea
\label{Creplace}
m_1 &\leftrightarrow& m_2\;,
\nonumber \\
m_{L_i}^2 &\leftrightarrow& (m_{L_i}^2+m_{e_j}^2)\;,
\nonumber \\
u&\rightarrow& d, s\ {\rm or}\ b\;.
\eea
For the bottom coupling the CCB-1 bound is
only valid if $\tan\beta\simlt 4$.

Concerning the CCB-2,3 bounds, they remain the same with the
previous (\ref{Creplace}) substitutions.
Moreover, for the estimation of $\hat Q_{d,s}$ we have to include
$\lambda_t\frac{A_{d,s}}{4\lambda_{d,s}}$
among the various quantities within the parenthesis above.
Now, there is no constraint on $\tan \beta$ and therefore possibility 4
in CCB-2,3
can be applied for any value of $\tan \beta$, disregarding possibility 5.

\vspace{0.3cm}
\noindent ${\bf \lambda_e L_e H_1 e_R}$,
${\bf\lambda_\mu L_\mu H_1 \mu_R}$, ${\bf\lambda_\tau L_\tau H_1 \tau_R}$

\vspace{0.3cm}
\noindent
The scale is given by:
$\hat Q_{e,\mu,\tau}\sim {\rm Max}\left(M_S, g_2\frac{A_{e,\mu,\tau}}
{4\lambda_{e,\mu,\tau}}\right)$.

The CCB bounds
remain the same with the following replacements
\bea
\label{Mreplace}
m_1 &\leftrightarrow& m_2\;,\hspace{2.3cm}
m_{L_i}^2 \rightarrow (m_{L_i'}^2+m_{e_j'}^2)\;,
\nonumber \\
u &\rightarrow& e,\mu\ {\rm or}\ \tau\;,\hspace{1.7cm}
(m_{L_i}^2+m_{e_j}^2) \rightarrow m_{L_i'}^2\;,
\nonumber \\
Q &\rightarrow& L\;.
\eea
where $L'_i$, $e'_{R_j}$ are two sleptons of a different generation
than the trilinear coupling under consideration.
When both extra sleptons appear in the bounds,
the associated Yukawa coupling, say $\lambda_l'$,
must be much smaller than the Yukawa coupling of the trilinear
coupling under consideration, say $\lambda_l$. Obviously this condition can
always be satisfied except when the coupling under consideration
is of the first generation (i.e. the electron one). In that case
$\alpha^2=1-\gamma^2$.

Here there is no constraint on $\tan \beta$ and therefore possibility 4
in CCB-2,3
can be applied for any value of $\tan \beta$ and possibility 5 should not
be taken into account. Moreover, for the estimation of
$\hat Q_{e,\mu,\tau}$ if we are considering
the CCB-2,3 bounds we have to include
$\lambda_t\frac{A_{e,\mu,\tau}}
{4\lambda_{e,\mu,\tau}}$
among the various quantities within the parenthesis above.

Under the assumption of universality
it is easy to see that the CCB-1 bound will only take place in the
possibility 1 [see condition above eq.(\ref{CCCB1a})],
while the CCB-2, CCB-3 bounds
will occur in the possibility 4 [note that the denominator of
eq.(\ref{CCCB2p}) goes to zero].

\vspace{0.3cm}
\noindent  ${\bf \lambda_t Q_t H_2t_R}$

\vspace{0.3cm}
\noindent
The CCB-1 bound does not apply to the top case.
Moreover, since  $\lambda_t=O(1)$ it is not true that a negative
minimum associated to the top trilinear coupling is necessarily
much deeper than the realistic minimum, thus destabilizing the
standard vacuum, as was the case of the previous couplings. Therefore, rather
than the absence of a negative minimum, we must demand that the
possible CCB minimum satisfies $V_{\rm CCB\;min}>V_{\rm real\;min}$.

When sign$(A_t)=-$sign$(B)$ (i.e. CCB-2), the potential is given by
\bea
\label{tValfabeta2}
V = \lambda_t^2 F'(\alpha) \alpha^4 |H_2|^4
 -2 \lambda_t \hat A'(\gamma) \alpha^2  |H_2|^3
 + {{\hat m}'^2}(\alpha,\gamma) |H_2|^2
- \frac{2 m_{L_i}^4}{g'^2+g_2^2}\;\;,
\eea
with
\bea
\label{tFfAm2}
F'(\alpha) &=&1+\frac{2}{\alpha^2} +\frac{f'}{\alpha^4}\;;\;\;f'=0\;,
\nonumber \\
\hat A'(\gamma)&=& |A_t|+|\mu|\gamma\;\;,
\nonumber \\
{{\hat m}'^2} (\alpha,\gamma) &=&
m_2^2+(m_{Q_t}^2+m_{t}^2)\alpha^2+m_1^2\gamma^2
+m_{L_i}^2(1-\alpha^2-\gamma^2)-2|m_3^2|\gamma.
\eea
This should be handled in the following way:

\begin{enumerate}

\item Scan $\gamma$ in the range $0\leq\gamma\leq 1$

\item For each value of $\gamma$ the optimum values of $\alpha^2$,
$H_2$
 i.e.
the ones that minimize the right hand side of (\ref{tValfabeta2}), are
given by
\bea
\label{talfext}
\alpha_{ext}^2=\frac{\hat A'(\gamma)}{\lambda_t |H_2|_{ext}}
-1 -\frac{m_{Q_t}^2+m_{t}^2-m_{L_i}^2}{2\lambda_t^2 |H_2|_{ext}^2}
\;\;,
\eea
\bea
\label{tH2ext}
|H_2|_{ext}=
\frac{3 \hat A'(\gamma)}{4\lambda_t\alpha_{ext}^2 F'(\alpha_{ext})}
\left\{1\ +\ \sqrt{1-\frac{8{{\hat m}'^2}(\alpha_{ext},\gamma)
F'(\alpha_{ext})}{9 \hat A'^2(\gamma)}}\;
\right\}\;\;.
\eea
For each value of $\gamma$ the coupled equations (\ref{talfext}),
(\ref{tH2ext}) can be solved, e.g. by a numerical method. Then,
the consistency of the procedure requires
\bea
\label{tconsist}
\alpha_{ext}^2>0,\;\;\;|H_2|_{ext}>0,\;\;\;
1-\gamma^2-\alpha_{ext}^2
-\frac{4 m_{L_i}^2}{(g'^2+g_2^2)|H_2|_{ext}^2}>0\;\;.
\eea
If (\ref{tconsist}) is fulfilled, then the corresponding value of the
potential at the minimum is given by
\bea
\label{tVCCBmin2}
V_{\rm CCB\ min}=-\frac{1}{2}\alpha_{ext}^2 |H_2|_{ext}^2
\left(\lambda_t \hat A'(\gamma) |H_2|_{ext}
-\frac{\hat m'^2(\alpha_{ext},\gamma)}{\alpha_{ext}^2}\right)
- \frac{2 m_{L_i}^4}{g'^2+g_2^2}\;\;.
\eea
Consequently, the CCB bound has the form
\be
\label{tCCBtop}
V_{\rm CCB\ min}(Q=\hat Q)> V_{\rm real\;min}(Q=M_S)
\;\;,
\ee
where $V_{\rm CCB\ min}$ and $V_{\rm real\;min}$ are given by
eqs.(\ref{tVCCBmin2}) and (\ref{Vreal}) respectively; the value of the
scale $M_S$ was explained in UFB-1 above and
$\hat Q\sim {\rm Max}\left(M_S, g_3\frac{A_t}{4\lambda_t},
\lambda_t\frac{A_t}{4\lambda_t}\right)$.

\item If (\ref{tconsist}) is not fulfilled,
then the potential is given by
\bea
\label{tValfabeta2b}
V = \lambda_t^2 F'(\alpha,\gamma) \alpha^4 |H_2|^4
 -2 \lambda_t \hat A'(\gamma) \alpha^2  |H_2|^3
 + {{\hat m}'^2}(\alpha,\gamma) |H_2|^2 \;\;,
\eea
The optimum values of $\alpha$, $H_2$ are now given by
\bea
\label{talfext2}
\alpha_{ext}^2=\frac{8\lambda_t^2}{g'^2+g_2^2+8\lambda_t^2}
\left[\frac{\hat A'(\gamma)}{\lambda_t|H_2|_{ext}}
-1-\frac{m_{Q_t}^2+m_{t}^2}{2 \lambda_t^2 |H_2|_{ext}^2}
+\frac{g'^2+g_2^2}{8 \lambda_t^2}(1-\gamma^2)\right]
\eea
\bea
\label{tH2ext2}
|H_2|_{ext}=
\frac{3 \hat A'(\gamma)}{4\lambda_t\alpha_{ext}^2 F'(\alpha_{ext},\gamma)}
\left\{1\ +\ \sqrt{1-
\frac{ 8 {\hat m}'^2 (\alpha_{ext},\gamma) F'(\alpha_{ext},\gamma) }
{ 9 \hat A'^2(\gamma) }     }\;\right\}\;\;,
\eea
with
\bea
\label{tFfAm3}
F'(\alpha,\gamma) &=&1+\frac{2}{\alpha^2} +\frac{f'}{\alpha^4}\;\;,
\nonumber \\
f'&=& \frac{g'^2+g_2^2}{8\lambda_t^2}
\left(1-\alpha^2-\gamma^2\right)^2\;\;,
\nonumber \\
{\hat m}'^2 (\alpha,\gamma) &=& m_2^2+\left(m_{Q_t}^2
+m_{t}^2\right)\alpha^2 + m_1^2\gamma^2 - 2|m_3^2|\gamma
\;\; .
\eea
Consistency now requires
\bea
\label{tconsist2}
\alpha_{ext}^2>0,\;\;\;|H_2|_{ext}>0\;\;.
\eea
Otherwise there is {\em no} CCB minimum for the particular value
of $\gamma$
being scanned. If (\ref{tconsist2}) is satisfied, then the value
of the potential at the minimum is given by
\bea
\label{tVCCBmin3}
V_{\rm CCB\ min} = -\frac{1}{2} \alpha_{ext}^2 |H_2|_{ext}^2
\left( \hat A'(\gamma) \lambda_t |H_2|_{ext}
- \frac{\hat m'^2(\alpha_{ext},\gamma)}{\alpha_{ext}^2} \right)\;\;.
\eea
and the CCB bound takes again the form
\be
\label{CCBtop2}
V_{\rm CCB\ min}(Q=\hat Q)> V_{\rm real\;min}(Q=M_S)
\;\;.
\ee

\end{enumerate}

When sign$(A_t)=$ sign$(B)$ (i.e. CCB-3) the analysis is exactly the same but,
as
usual,  one of the three terms proportional to
$|A_t|,\ |\mu|\gamma,\ |m_3^2|\gamma$ in eqs.(\ref{tFfAm2}),
(\ref{tFfAm3}) must flip its sign.

Let us finally note that if $\tan\beta$ is large ($\tan\beta\simgt
15$), then $\lambda_b=O(1)$ and the analysis of this subsection
is also the correct one for the bottom, performing the substitutions
\bea
\label{treplace2}
H_1 &\leftrightarrow& H_2\;, \hspace{2.35cm}m_{L_i}^2\rightarrow
(m_{L_i}^2+m_{e_j}^2)\;,
\nonumber \\
m_1 &\leftrightarrow& m_2\;,\hspace{3.1cm}t\rightarrow b\;.
\eea

\section{Constraints on the parameter space}

In the previous sections, a complete analysis of all the potentially
dangerous unbounded from below (UFB)
and charge and color breaking (CCB)
directions has been carried out. In particular, the analytical form of the
constraints obtained on the parameter space of the MSSM
has been summarized in sect.5.
Now, we will analyze numerically those constraints. We will see that they
are very important and, in fact,
there are {\it extensive regions} in the parameter space which are forbidden.

Our analysis will be quite general in the sense that we will consider the
whole parameter space of the MSSM, $m$, $M$, $A$, $B$, $\mu$, with the only
assumption of universality\footnote{Let us
remark, however, that the constraints
found in previous sections are general and they could also be applied
for the non-universal case.}. Actually, universality
of the soft SUSY-breaking terms at $M_X$ is a
desirable property not only to reduce the number of
independent parameters, but also for phenomenological reasons, particularly
to avoid flavour-changing neutral currents (see, e.g. ref.\cite{Ross}).
As discussed in sect.2, the requirement of correct electroweak breaking
fixes one
of the five independent parameters of the MSSM, say $\mu$, so we are left
with only four parameters ($m$, $M$, $A$, $B$). Although we will perform
the numerical analysis on this space, it is worth noticing that
particularly interesting values of $B$ can be obtained from
Supergravity (SUGRA). In this sense we will first consider two values of
$B$ as guiding examples to get an idea of
how strong the different constraints are
and then we will vary $B$ in order to obtain the most
general results. Hence, let us first justify, theoretically and
phenomenologically, the two specific values of $B$.


The particular values of the soft terms depend on the type of Supergravity
theory from which the MSSM derives and, in general, on the mechanism of
SUSY-breaking. But, in fact, is still possible to learn things about soft
terms without knowing the details of
SUSY-breaking \cite{Yo}. Let us consider the simple
case\footnote{We will assume from now on a vanishing cosmological constant.}
of canonical kinetic terms for hidden and observable matter fields (i.e. a
K\"ahler potential $K=\sum_\alpha |\phi_\alpha|^2$). Then,
irrespective of the SUSY-breaking mechanism, the scalar masses are
automatically universal. Furthermore, if the observable part of the
superpotential $W$ is
assumed to be as in eq.(\ref{W}), $\mu$ being an initial parameter, then
the $B$ term and the universal
$A$ terms are automatically
generated and they are related to each other (assuming
that Yukawa couplings and $\mu$ are hidden field independent \cite{Yo}) by the
well known relation \cite{Weinberg}
\be
\label{BAm}
B = A - m \ .
\ee
Finally, if the gauge kinetic function is the same for the different gauge
groups of the theory $f_a=f$ (where $a$ is associated with $SU(3)$, $SU(2)_L$
and $U(1)_Y$), the gaugino masses are also universal.
This SUGRA theory is attractive for its simplicity and for the natural
explanation that it offers to the universality of the soft terms.
However, this scenario has a serious drawback. It is well known that, in
order to get appropriate $SU(2)_L\times U(1)_Y$ breaking, the $\mu$ parameter
has to be of the same order of magnitude ($M_W$) as the soft SUSY-breaking
terms discussed above. This is in general unexpected since the $\mu$ term
is a SUSY term whereas the soft terms are originated after SUSY-breaking.
In principle, the natural scale of $\mu$ would be the Planck mass.
The unnatural smalleness of the $\mu$ parameter is the so-called
$\mu$ problem. We will briefly explain here three interesting
scenarios considered in SUGRA in order to solve the problem,
illustrating them in the case
of canonical kinetic terms:

{\em (a)} In ref.\cite{Casas} was pointed out that the presence of a
non-renormalizable term in the
superpotential, $\lambda W H_1 H_2$, characterized by the coupling $\lambda$,
yields dynamically a
$\mu$ parameter when the hidden sector part of $W$ acquires a VEV, namely
$\mu = m_{3/2} \lambda $, where $m_{3/2}$ is the gravitino mass.
The fact that $\mu$ is of the electroweak scale order is a consequence
of our assumption of a correct SUSY-breaking scale
$ m_{3/2} = O(M_W) $. Now, with this solution to the $\mu$ problem, the
$B$ parameter can be straightforwardly evaluated.
The simple result (in the case of
$\lambda$ independent of the hidden fields \cite{Yo}) is
\be
\label{B2m}
B = 2 m \ .
\ee
For this mechanism to work, the $\mu H_1 H_2$ term in eq.(\ref{W})
must be initially absent (otherwise the natural scale for $\mu$ would be the
Planck
mass), a fact that remarkably enough, is automatically guaranteed in the
framework of Superstring theory as we will see below.

{\em (b)} In refs.\cite{Masiero, Casas}
it was shown that if a term, $Z H_1 H_2 + h.c.$,
characterized by the coupling $Z$
is present in the K\"ahler potential, an effective low-energy $B$ term is
naturally generated.
In the case of $Z$ independent of the hidden fields, this mechanism
for solving the $\mu$ problem is equivalent \cite{Casas} to the previous
one {\em (a)} and therefore
the value of $B$ is again given by eq.(\ref{B2m}). Now, the size
of $\mu$ is $\mu = m_{3/2} Z$.

{\em (c)} In ref.\cite{Giudice} the observation was made that in the framework
of
any SUSY-GUT, starting again with $\mu = 0$, an effective $\mu$ term is
generated by the integration of the heavy degrees of freedom.
The prediction for $B$ is once more given by eq.(\ref{B2m}).

The solutions discussed here in order to solve the $\mu$ problem
are {\it naturally present} in Superstring theory. In ref.\cite{Casas}
was first remarked that the $\mu H_1 H_2$ term is naturally absent
as already mentioned above. The reason is that in SUGRA theories coming
from Superstring theory mass terms for light fields are forbidden in the
superpotential.
Then a realistic example where non-perturbative
SUSY-breaking mechanisms like gaugino-squark condensation induce
superpotentials of the type {\em (a)} was given. In ref.\cite{Narain}
the same kind of superpotential was obtained using pure gaugino condensation
in the context of orbifold models.
The alternative mechanism {\em (b)} in which there is an extra term in the
K\"ahler
potential originating a $\mu$-term
is also naturally present in some large classes of four-dimensional
Superstrings \cite{Kaplu,Lust,Narain}. In Superstring theory, neither the
kinetic terms are in general canonical nor the couplings
(Yukawas, $\lambda$, $Z$) and the mass term ($\mu$)
are independent of hidden fields.
However, it is still possible to obtain (the phenomenologically
desirable) universal soft terms in the so-called
dilaton-dominated limit \cite{Kaplu,Brignole}. This limit is not only
interesting because of that, but also because
it is quite {\it model independent}
(i.e. for any compactification scheme the results for the soft terms are the
same).
It is also remarkable, that in this
limit once again the value of $B$ for the two mechanisms {\em (a), (b)}
coincides \cite{Yo} with that of eq.(\ref{B2m}).
If, alternatively, we just assume that a small ($\sim M_W$) dilaton-independent
mass $\mu$
is present in the superpotential, then the result for $B$ is
now given \cite{Brignole} by eq.(\ref{BAm}) as in the case of canonical
kinetic terms.

{}From the above analysis, it is clear that eqs.(\ref{BAm},\ref{B2m})
give us two values of $B$ very interesting from the theoretical and
phenomenological point of view. Thus, we will consider, for the moment, in
our numerical study of the UFB and CCB constraints both possibilities.
In fact, the value
of $\mu$ is also fixed once we choose a particular mechanism for
solving the $\mu$ problem, e.g. mechanisms {\em (a), (b)} (see above).
However, this value still depends on the couplings
$\lambda$ and $Z$
which are in general model dependent\footnote{
For an
analysis of the MSSM from Superstring theory taking into account a particular
value of $Z$ coming from
orbifold compactifications, and therefore a fixed value of
$\mu$, see ref.\cite{BIMS}.}, so we prefer to eliminate $\mu$ in terms of
the other parameters by imposing appropriate symmetry-breaking
at the weak scale as mentioned above.
Let us now turn to the numerical results.

In Fig.1 we have presented in detail the case $B=A-m$ with $m=100$ GeV, to get
an idea of how strong the different constraints are, plotting the excluded
regions in the  remaining
parameter space ($A/m$, $M/m$). It is worth noticing here that
even before imposing CCB and UFB constraints, the parameter space is strongly
restricted by the experiment. As already mentioned in sect.2,
not for all the parameter space
it is possible to choose the boundary condition of $\lambda_{top}$
so that the experimental mass of the top is reproduced, since the RG infrared
fixed point of $\lambda_{top}$ puts an upper bound on $M_{\rm top}$,
namely $M_{\rm top}\simlt 197\sin\beta$ GeV \cite{infrared},
where $\tan\beta=v_2/v_1$.
In this way, the upper and lower darked regions are forbidden because
$M^{\rm phys}_{\rm top}=174$ GeV cannot be reached.
Furthermore, the small central darked region is also forbidden because
there is no value of $\mu$ capable of producing the correct electroweak
breaking.

Fig.1a shows the region excluded by  the ``traditional" CCB
bounds of the type of
eq.(\ref{frerebound}), evaluated at an {\em appropriate} scale
(see subsect.4.5). For a point in the parameter
space to be excluded we have also demanded that the corresponding
CCB minimum is deeper than the realistic one (this is especially
relevant for the bounds coming from the top trilinear term).
Clearly, the ``traditional" bounds, when correctly evaluated, turn out to
be very weak. In fact, only the leptonic (circles) and the $d$--type
(diamonds) terms do restrict, very modestly, the parameter space.
Let us recall here that it has been a common (incorrect) practice
in the literature to evaluate these traditional bounds at all the scales
between $M_X$ and $M_W$, thus obtaining very important (and of course
overestimated) restrictions in the parameter space.
Fig.1b shows the region excluded by our ``improved" CCB constraints
obtained in sect.4 and summarized in sect.5. Comparing Figs.1a
and 1b it is clear that the excluded region becomes dramatically increased.
Notice also that all the trilinear couplings (except the top one in this
case) give restrictions, producing areas constrained by different
types of bounds simultaneously. The restrictions coming from the UFB
constraints, obtained in sect.3 and summarized in sect.5, are shown
in Fig.1c. By far, the most restrictive bound is the UFB--3 one (small
filled squares). Indeed, the UFB--3 constraint is the {\em strongest}
one of {\em all} the UFB and CCB constraints, excluding extensive
areas of the parameter space, as is illustrated in the figure. In our
opinion, this is a most remarkable result.
Finally, in Fig.1d we summarize all the constraints
plotting also the excluded region due to the (conservative)
experimental bounds on SUSY
particle masses (filled diamonds) of eq.(\ref{Expb}).  More precisely,
this forbidden area comes from too
small masses for the gluino, lightest chargino, lightest
neutralino, left sbottom, and left and right $u,c$ squarks.
The allowed region left at the end of the day
(white) is quite small.

Figs.2a, 2b, 2c give, in a summarized way, the same analysis as that
of Fig.1, but for three different values of $m$ ($m=100$ GeV,
$m=300$ GeV, $m=500$ GeV).
For the plots with $m$ bigger
than 100 GeV the gluino, lightest stop, lightest chargino and
lightest neutralino are responsible for
the excluded region due to experimental
bounds on masses.
The ants indicate regions
which are excluded by negative squared
mass eigenvalues, in this case
the lightest stop.
The figures show a clear trend in the sense that
the smaller the value of $m$, the more restrictive the
constraints become.
This is mainly due to
the effect of the UFB-3 constraint (note the almost exact $m$--invariance
of the CCB bounds).
In the limiting case $m=0$ (not represented in the
figures) essentially the
{\it whole} parameter space turns out to be excluded. This has obvious
implications, e.g. for no-scale models\footnote{We thank J. L\'opez
for a comment stressing us the possible implications of the CCB and UFB
bounds for no-scale models.} \cite{noscale}.
Anyway,
extensive areas in the parameter space are forbidden in all cases.

The same conclusions are obtained for the other (theoretically and
phenomenologically well-motivated) value of $B$, $B=2m$. The
results in this scenario are shown
in Fig.3, where the whole darked region is forbidden because
$M^{\rm phys}_{\rm top}=174$ GeV cannot be reached. Unlike the Fig.2,
now in some cases the left sbottom
may also get a negative squared mass eigenvalue.

Finally, in Figs.4a, 4b we generalize the previous analyses by
varying the value
of $B$ for different values of $m$, namely $m=100$ GeV,
$m=300$ GeV.
The final allowed regions from all types of bounds
in the parameter space of the
MSSM are shown.
Both figures exhibit a similar trend. For a particular value of $m$,
the larger the value of
$B$ the smaller the allowed region becomes. More precisely, the maximum
allowed value of $B$ is $B=2.5m$ for $m=100$ GeV and $B=3.5m$
for $m=300$ GeV. This fact comes mainly
from the enhancement of the forbidden areas by the UFB-3 constraint and the
requirement of $M^{\rm phys}_{\rm top}=174$ GeV.
Both facts are due to the decreasing of $\tan \beta$ as $B$ grows.
Then higher top Yukawa couplings are needed in order to reproduce
the experimental top mass. On the one hand, this cannot be always
accomplished due to the infrared fixed point limit on the top
mass. On the other hand, the larger the top Yukawa coupling, the
stronger the UFB-3 bound becomes.
For negative values of $B$
the corresponding figures can easily be deduced from the previous ones,
taking into account that they are invariant
under  the transformation
$B,A,M \rightarrow -B,-A,-M$.

{}From the various figures it is clear that the CCB and UFB constraints
{\em put
important bounds} not only on the value of $A$, but also on the values of
$B$ and $M$, which is an interesting novel fact.

\section{Conclusions}

Although the possible existence of dangerous charge and color breaking
minima in the supersymmetric standard model has been known since the
early 80's, a complete study of this crucial issue was still lacking. This
was due to two reasons: First, the complexity of the SUSY scalar potential,
$V$,
caused that only particular directions in the field-space were considered, thus
obtaining necessary but not sufficient conditions to avoid dangerous
charge and color breaking minima. Second, the radiative corrections to $V$
were not normally included in a proper way.

In the present paper we have carried out a
complete analysis of all the potentially
dangerous directions in the field-space of the MSSM, obtaining the
corresponding constraints on the parameter space. These are completely general
and can be applied to the non-universal case. The
constraints turn out to be very important and, in fact, there are
{\it extensive regions} in the parameter space which are forbidden, increasing
the predictive power of the theory.

The constraints can be clasified in two types. First, the ones
associated with the existence of charge and color breaking (CCB) minima in
the potential deeper than the realistic minimum.
Second, the constraints associated with
directions in the field-space along which the potential becomes
unbounded from below (UFB). It is worth mentioning here that the unboundedness
is only true at tree-level since radiative corrections eventually
raise the potential for large enough values of the fields, but still
these minima can be deeper than the realistic one and thus dangerous.

We have performed a complete analysis of both types of directions
obtaining {\it new and very restrictive bounds}, expressed in an analytic
way, that represent a set of necessary and sufficient constraints.
They are summarized in sect.5.
For certain values of the initial parameters the CCB constraints
``degenerate" into the UFB constraints since the minima become unbounded from
below directions. In this sense, the CCB constraints comprise the UFB bounds,
which can be considered as special (but extremely important) limits of the
former.

We have also taken into
account the radiative corrections to $V$ in a proper way. To this respect,
let us remember that, usually, the scalar potential
is considered at tree-level, improved by
one-loop RGEs, so that all the parameters appearing in it are running
with the renormalization scale, $Q$. Then it is often demanded that the
CCB and UFB constraints are satisfied at any scale between
$M_X$ and $M_Z$. However, this is not correct since the tree-level scalar
potential is strongly $Q$-dependent and the one-loop
radiative corrections to it are crucial to make the potential stable against
variations of the scale. Using the scale independence of $V$, instead
of minimizing the complete one-loop potential, which would be an impossible
task, we have demanded
that the previous (tree-level-like) bounds are satisfied at the renormalization
scale, $Q$, at which the one-loop correction to the
potential is essentially negligible. This simplifies enormously
the analysis, producing equivalent results. We have also given explicit
expressions of the appropriate scale to evaluate the different types of bounds.

The usual lack in the literature
of an optimum scale to evaluate the constraints
implies that their restrictive power has normally been overestimated.
E.g., the ``traditional" CCB bounds (see eq.(\ref{frerebound}))
when (incorrectly) analyzed
at $M_X$ are very strong. However, we have seen that
when correctly evaluated, they turn out to
be very weak (see Fig.1a).
The new CCB constraints obtained here are much more restrictive and, in fact,
the excluded region becomes dramatically increased (see Fig.1b). On the
other hand, the restrictions coming from the new UFB constraints are by
far the most important ones, excluding extensive areas of the parameter
space (see e.g. Fig.1c).

We have performed a numerical analysis of how our UFB and CCB constraints
put restrictions on the whole parameter space of the MSSM. As already
mentioned they are very strong producing important bounds not only on the
value of $A$ (soft trilinear parameter), but also on the values of $B$
(soft bilinear parameter) and $M$ (gaugino masses). This is a new and
interesting feature. This analysis is summarized in Figs.2--4.
As a general trend, the smaller the value of $m$, the more restrictive the
constraints become. In the limiting case $m=0$ essentially the
{\it whole} parameter space turns out to be excluded. This has obvious
implications, e.g. for no-scale models.

Finally, let us mention that all the constraints that has been obtained here
come from the requirement that the standard vacuum is the global minimum of
the theory.
Although the possibility of living in a metastable
vacuum with a lifetime larger than the present age of the Universe
\cite{Claudson} does not seem specially attractive, it cannot be excluded.
Since the constraints on the parameter space found in this paper are very
strong, this dynamical question deserves further analysis
\cite{Nosotros}.



\newpage

\section*{Appendix}

In subsect.4.1 we have enumerated five general properties concerning charge and
color breaking (CCB) minima in the MSSM. Properties 1, 3, 5 remained to
be proved, which is the aim of this appendix.

Let us however first notice that some of the properties (in particular
the 1 and 3 ones) can be intuitively understood by a simple consideration.
Suppose we consider a region in the field-space where only one trilinear
scalar term is non-negligible. Denoting by $\phi$ the typical size of
the relevant VEVs at a CCB minimum, we can schematically write the relevant
terms in the potential as
\be
\label{intui}
V\sim Nm^2\phi^2-2A\lambda \phi^3 + N'\lambda^2\phi^4\ +\ {\rm D-terms}
\;\;,
\ee
where $N,N'= O(1)$ (typically $N,N'\sim 3$), $m,A\sim M_{S}$ (i.e. the
scale of SUSY breaking) and $\lambda$ is the Yukawa coupling (note that
with a convenient  choice of the field phases, the trilinear scalar term
can always be made negative as in (\ref{intui})). Ignoring for the moment
the D--terms, it is clear that $V$ will only be negative in the range
\be
\label{intui2}
\frac{Nm^2}{2A\lambda}< \phi < \frac{2A}{N'\lambda}
\;\;,
\ee
which implies
\be
\label{intui3}
\phi\sim\frac{M_S}{\lambda}
\;\;.
\ee
Now, if $\lambda\ll 1$, then $\phi\gg M_S$. In that case it is clear that
the D--terms must be essentially cancelled (i.e. property 3), otherwise they
would contribute a positive amount of order $g^2|\phi|^4$ that would
dominate the potential (\ref{intui}). Furthermore, from (\ref{intui3}),
it follows that two trilinear scalar terms with different Yukawa couplings
cannot efficiently "cooperate" to improve the CCB bounds (i.e. property 1):
the potential
can only be negative in two separate regions in the field-space given by
eq.(\ref{intui3}) applied to each coupling. In any of these regions, the
presence of the extra trilinear term plus the associated mass and F terms
can only yield a positive contribution to the potential. An explicite
example of this argument can be found below eq.(\ref{H1Fterm})

\subsection*{Property 1}

As we have already mentioned, according to this property the most dangerous
CCB directions in the MSSM potential involve only one particular trilinear
soft term.

Since it is not possible to get an analytical formulation of the general
CCB minima with all the fields and couplings in the game, the proof of the
previous statement can only come from an exhaustive analysis of all
the ways in which two or more different trilinear scalar terms could
cooperate to improve the CCB bounds. Next we consider all the cases
in a separate way\footnote{We simplify somewhat the notation (in an
obvious way) to go more straightforwardly through the arguments.
Likewise, in some specific points we will use the
assumption of universality to simplify the arguments, but these can easily be
extended with slight modifications to more general cases.}.

\vspace{0.3cm}
\noindent ${\bf \lambda  Q H_2u}+{\bf \lambda' Q'H_2u'}$ ;
${\bf \lambda\ll \lambda'}$

\vspace{0.2cm}
\noindent
Here we consider the simultaneous presence in the Lagrangian of two
different couplings of the $u$ type and the corresponding terms in
the scalar potential from the associated D--terms, F--terms and soft
terms. According to the notation of the heading, the pair of quarks
$\{u,u'\}$ may represent $\{u,c\}$, $\{u,t\}$ or $\{c,t\}$. It is
convenient for our analysis to roughly divide the field-space in the three
following regions

\begin{description}

\item[a)] $Qu\ll Q'u'$

\item[b)] $Qu\sim Q'u'$

\item[c)] $Qu\gg Q'u'$

\end{description}
\noindent
where $Q,u,Q',u'>0$ without loss of generality. Let us examine the
CCB issue in each zone separately, taking for simplicity $Q=u$,  $Q'=u'$.

\begin{description}

\item[a)]

All terms in the potential involving $Q$ and/or $u$ are
negligible, so the only significant term is the $\lambda'$ one.
Therefore the $(a)$ area is irrelevant for property 1.

\item[b)]

In this region $A\lambda  Q H_2u\ll A'\lambda' Q'H_2u'$, so,
again, property 1 cannot be disproved here. We can check however that
the region $(b)$ is anyway irrelevant for CCB bounds. The only terms
in $V$ where the presence of $Q,u$ is relevant are
\be
\label{b_terms}
(m_Q^2 + m_u^2) Q^2\ +\ {\rm D-terms}
\;\;,
\ee
where we have used $Q=u$.

In the case where $\lambda=\lambda_u$, $\lambda'=\lambda_c$, it
happens that, very accurately, $m_Q^2 = m_{Q'}^2$,
$m_u^2 = m_{u'}^2$. Therefore $Q^2$ occurs in $V$ only through
the combination $\hat Q^2\equiv Q^2+Q'^2$. Along any direction
with $Q^2/Q'^2=$const. the relevant terms in the potential can
be writen as
\be
\label{b_terms2}
-2A'\hat\lambda H_2 \hat Q^2 +(m_Q^2 + m_u^2) \hat Q^2
+\hat \lambda^2\hat Q^4 + 2\hat \lambda^2 H_2^2 \hat Q^2
(1+\frac{Q^2}{Q'^2}) + {\rm D-terms} + \cdots
\;\;,
\ee
where $\hat\lambda=\lambda' Q'^2/\hat Q^2$ and the D--terms are a function
of $\hat Q^2$. Therefore everything occurs as if there were a single
coupling $\hat \lambda\hat Q H_2 \hat u$, except for the additional
(positive) term proportional to $\frac{Q^2}{Q'^2}$. Recalling now that
in the case of a coupling $\ll 1$, the general CCB bound does not depend
on the value of the coupling itself\footnote{For more details, see
sect.4, e.g. eqs.(\ref{GenCond}), (\ref{GenConda}).}, it is clear that the
optimum
direction arises for $\frac{Q^2}{Q'^2}=0$. Thus the $(b)$ region
is irrelevant.

When $\lambda'=\lambda_t$, the previous argument is not valid, but
it is still true from (\ref{b_terms}) that the same role of $Q$
can be played by a slepton
$L$ with exactly the same VEV along the $\nu_L$ direction.
Then the D--terms are
exactly the same but, since $m_L^2<m_Q^2+m_u^2$, it is clear that
the potential becomes deeper. Consequently, the $(b)$ region does
never correspond to an improved CCB bound.

\item[c)]

In this region the effect of $Q'$, $u'$ in the mass terms is
negligible and it is convenient to look at the potential ``from the
point of view" of $\lambda Q H_2 u$ as the relevant coupling. The
relevant terms of the potential are
\bea
\label{Vc}
V &=& \left(m_{Q}^2 +m_{u}^2\right) Q^2
- 2AH_2\left(\lambda Qu + \lambda'Q'u'\right)
+m_2^2H_2^2+m_1^2H_1^2-2m_3^2H_1H_2
\nonumber \\
&+&
\left|\mu H_1+\lambda Qu + \lambda'Q'u'\right|^2
+ 2\left|\lambda H_2Q\right|^2
+ 2\left|\lambda' H_2Q'\right|^2
\nonumber \\
&+&
\frac{1}{8}(g^2+g'^2)\left|H_2^2-H_1^2-Q^2\right|^2
\;\; ,
\eea
where for simplicity we have taken $A=A'$.

For $\lambda'Q'\sim\lambda Q$, or smaller, it is clear that the only
non-negligible term involving $Q'$ is $2|\lambda'H_2Q'|^2$, which is
positive. Thus, a value of $Q'$ of this order can never be useful to
make the potential deeper.

For greater values of $Q'$, in particular $\lambda'Q'u'\sim\lambda Q u$,
there appear new relevant terms in the potential involving $Q',u'$, as
can be seen from (\ref{Vc}). In this case the potential (\ref{Vc})
can be reformulated as if it was derived from a single coupling
$\hat \lambda QH_2u$ with $\hat\lambda\equiv \lambda(1\ +\
\lambda'Q'^2/\lambda Q^2)$, except for the terms
\bea
\label{doslam}
2\left|\lambda H_2Q\right|^2
+ 2\left|\lambda' H_2Q'\right|^2=
2\left|\hat\lambda H_2Q\right|^2\frac
{1+\frac{\lambda'Q'^2}{\lambda Q^2}\frac{\lambda'}{\lambda}}
{\left(1+\frac{\lambda'Q'^2}{\lambda Q^2}\right)^2}\;\;,
\eea
which appear instead of the $2\left|\hat\lambda H_2Q\right|^2$ term.
Since $\lambda'\gg\lambda$, it is clear that as long as
$\frac{\lambda'Q'^2}{\lambda Q^2}\ll\frac{\lambda'}{\lambda}$
(which, by definition, always occurs in the $(c)$ region),
(\ref{doslam}) is bigger than $2\left|\hat\lambda H_2Q\right|^2$,
so the CCB bounds obtained in this region are less stringent
than those obtained by consideration of a unique coupling
$\lambda QH_2u$ (recall that for small couplings the form of the
CCB bound does not depend on the value of the coupling itself).

\end{description}

\vspace{0.3cm}
\noindent ${\bf \lambda  Q H_1d}+{\bf \lambda' Q'H_1d'}$ ;
${\bf \lambda\ll \lambda'}$

\vspace{0.2cm}
\noindent
This case can be analyzed along similar lines than the previous
heading, with analogous results.

\vspace{0.3cm}
\noindent ${\bf \lambda  Q H_1d}+{\bf \lambda' Q'H_2u'}$ ;
${\bf \lambda\ll \lambda'}$

\vspace{0.2cm}
\noindent
Again, we divide the field-space in the three regions

\begin{description}

\item[a)] $Qd\ll Q'u'$

\item[b)] $Qd\sim Q'u'$

\item[c)] $Qd\gg Q'u'$

\end{description}
\noindent
with $Q,d,Q',u'>0$ .

\begin{description}

\item[a)]

Similarly to the previous heading, this case is irrelevant.

\item[b)]

In this region the trilinear scalar term $\lambda A QH_1d$
is negligible, so property 1 cannot be disproved. Let us note anyway
that the only relevant terms involving the $Q,d$ fields are the mass
terms and the D--terms. Hence their role can be more profitably played
by sleptons, which have lower masses. More precisely, a single slepton
$L$ (taken along the $\nu_L$ direction) will be needed if
$H_2^2-Q'^2-H_1^2>0$, while two sleptons, $L$ (along $l_L$), $l_R$,
will be needed if $H_2^2-Q'^2-H_1^2<0$. (Furthermore the leptonic
coupling $\lambda_l$ must be $\lambda_l\ll\lambda'$, so a choice
that always works is to take the slepton from the first generation.)

\item[c)]

Analogously to the previous heading, this region is more
conveniently seen ``from the point of view" of $\lambda QH_1d$ as
the relevant coupling. The only relevant terms in the potential
involving $Q',u'$ are
\bea
\label{tres_c}
-2A'\lambda' Q'^2H_2
+\left|\mu H_1 + \lambda' Q'^2\right|^2
+ 2\left|\lambda' H_2Q'\right|^2
\eea
(recall we are taking $Q'=u'$). Clearly, if $Q'\neq 0$, only the
first two terms can be useful to make the potential deeper. The first
term will only be significant if $H_2\sim H_1$, but then the (positive)
third term dominates the potential. Therefore we conclude that
$Q'\neq0$ can only be relevant for the CCB bounds if $H_2=0$ or
negligible. Then, the $Q'$ value can be optimally adjusted so that
\bea
\label{uno_c}
\left|\mu H_1 + \lambda' Q'^2 \right|^2 = 0\;\;.
\eea
Of course, this possibility has been considered in the analysis of
the CCB bounds (see CCB--1 bound in the main text). In any case, note
that, since $H_2=0$, the {\em only} relevant trilinear scalar term
is $\lambda QH_2d$, in agreement with property 1.

\end{description}

\vspace{0.3cm}
\noindent ${\bf \lambda  Q H_1d}+{\bf \lambda' Q'H_2u'}$ ;
${\bf \lambda\gg \lambda'}$

\vspace{0.2cm}
\noindent
This case is completely analogous to the previous one interchanging
$Q\leftrightarrow Q'$, $d\leftrightarrow u'$, $H_1\leftrightarrow H_2$,
$\lambda\leftrightarrow \lambda'$,

\vspace{0.3cm}
\noindent ${\bf \lambda  L H_1l}\ +$ {\bf other couplings}

\vspace{0.2cm}
\noindent
The analysis is completely similar to that of
$\lambda  Q H_1d\ +$ other couplings in the three previous headings.
The only exception is that when $\lambda L H_1 l$ corresponds to
the electron coupling there is no slepton, say $L'$, with smaller
Yukawa coupling, to play with (the existence of such an slepton
is used when analyzing the $(b)$ region above). However, this is
irrelevant in practice since

\begin{itemize}

\item The leptonic couplings of $\mu,\tau$ turn out to give more
stringent CCB restrictions than the electron one, as can be seen
in the text (see sect.6).

\item For all the leptonic couplings the direction with $L'\neq 0$
is never the most dangerous one.

\end{itemize}

\vspace{0.3cm}
\noindent {\bf Two couplings with}
${\bf \lambda\sim \lambda'}$

\vspace{0.2cm}
\noindent
This case represents the only possible exception to property 1.

A paradigmatic example would be to consider the bottom and tau
couplings
\bea
\label{btau}
\lambda_b QH_1b + \lambda_\tau LH_1\tau\;\;.
\eea
In the extreme (and non--realistic) case that $\lambda_b=\lambda_\tau$,
$m_Q^2=m_L^2$, $m_b^2=m_\tau^2$, $A_b=A_\tau$ (at the correct scale),
it is easy to see that for a given value of $|\phi|^2\equiv
|Q|^2 + |L|^2$ the potential is independent of the particular
values of $|Q|^2, |L|^2$. In practice, however, the previous
equalities do not hold, in particular, typically
$m_Q^2>m_L^2$, $m_b^2>m_\tau^2$. Hence, it becomes profitable
to use just one of the two VEVs, typically $|L|^2$,
as it is confirmed by the numerical results (see sect.6).
Consequently, in this case property 1 holds.

Other examples can arise for particular values of $\tan\beta$.
For example
\bea
\label{ul}
\lambda_u QH_2u + \lambda_l LH_1l\;\;.
\eea
can have $\lambda_u\sim\lambda_l$ for particular choices
of ($u,l$) and particular values of $\tan\beta$ (e.g.
for ($u,l$) $=$ ($c,\tau$) and $\tan\beta\sim2$).

Playing just with the fields appearing in (\ref{ul}), it is
possible to arrive to an optimized CCB condition
\bea
\label{ulcond}
&&\left(|A_u|+ |A_l|\gamma\frac{\lambda_l}{\lambda_u}
\frac{\gamma_l^2}{\alpha^2} + |\mu|\gamma
+|\mu|\frac{\lambda_l}{\lambda_u}\frac{\gamma_l^2}{\alpha^2}\right)^2\
\nonumber \\
&<& \left[1+\frac{2}{\alpha^2}
+\left(\frac{\lambda_l}{\lambda_u}\right)^2
\left(\left(\frac{\gamma_l}{\alpha}\right)^4
+ 2 \frac{\gamma^2\gamma_l^2}{\alpha^4}\right)
\right]
\nonumber \\
&\times&\left[ m_2^2+(m_{Q}^2+m_{u}^2)\alpha^2
+ m_1^2\gamma^2 + (m_{L}^2+m_l^2)\gamma_l^2 - 2|m_3^2|\gamma
\right]
\eea
where $\gamma=H_2/H_1$, $\alpha^2=Q^2/H_2^2$, $\gamma_l^2=
L^2/H_2^2$ and $1-\gamma^2-\gamma_l^2-\alpha^2=0$.
Actually, eq.(\ref{ulcond}) holds if sign$(A)$=--sign$(B)$. In the opposite
case we have to change the sign either of the $\propto |m_3^2|$ or
of the $\propto |\mu|$ terms in the previous equation.
We have not used this type of condition in the examination
of the CCB bounds of the MSSM (see sections 4--6).

\subsection*{Property 3}

In the general property 3 of subsect.4.1 it was stated that
if the trilinear term under consideration has a Yukawa coupling
$\lambda^2\ll 1$, which occurs in all the cases except for the top,
then the corresponding deepest CCB direction occurs for
vanishing (or negligible) D--terms. Next, we prove this property taking
for definiteness the trilinear coupling
\be
\label{guidex}
\lambda QH_2u
\ee
as the relevant coupling and considering (a priori) non-vanishing VEVs
for the fields $H_2$,$Q$ (taken along the $u_L$ direction), $u$, $H_1$,
parameterized as
\be
\label{fieldsrel}
|Q|=\alpha |H_2|,\;\;|u|=\beta |H_2|,\;\;|H_1|=\gamma |H_2|
\ee
(a non-vanishing VEV for a slepton could be also included in the analysis).
For simplicity we will focuss on the $SU(2)\times U(1)$ D--terms, so we will
assume for the moment
\be
\label{su3}
\alpha=\beta
\ee
and thus $Q=u$. Then the corresponding scalar potential has the form
\bea
\label{Vd1}
V &=& \left(m_{Q}^2 +m_{u}^2\right) |Q|^2
+m_2^2|H_2|^2+m_1^2|H_1|^2 - \left(m_3^2H_1H_2\ +\ {\rm h.c.}\right)
\nonumber \\
&+&
2\left|\lambda H_2Q\right|^2
+ \left|\lambda Q^2\right|^2
\nonumber \\
&+&
\left(\lambda AQ^2H_2 + \lambda\mu Q^2H_1^*\ +\ {\rm h.c.}\right)
\nonumber \\
&+&
\frac{1}{8}(g^2+g'^2)\left[|H_2|^2-|H_1|^2-|Q|^2\right]^2
\;\; .
\eea
The strategy of our proof is to suppose that the values of the fields
are in such a way that D--terms$\neq 0$, and then show that no CCB
minimum can arise in this situation.

The first consideration is that if D--terms$\neq 0$, then, necessarily,
all the terms involving $\lambda$ are irrelevant. For the quartic F--terms
(second line of (\ref{Vd1})), this is obvious since $\lambda^2\ll
(g^2+g'^2)$. The trilinear terms (third line of (\ref{Vd1})) can only
be competitive with the D--terms if the generic value of the involved
fields (say $\phi$) is $|\phi|\simlt \frac{\lambda}{g^2}M_S$ (recall that
$A,\mu=O(M_S)$). In that case, both the trilinear and the D--terms
are negligible compared to the mass terms. Let us also note that if the
values of the fields are tuned in such a way that the D--terms are
non-vanishing but small enough to be comparable with the rest of the terms,
then it is always favoured to slightly modify those values so that
D--terms$\rightarrow 0$ (or negligible), since this is accomplished
with almost no cost in the rest of the terms. Consequently, in any case
we can write the scalar potential as
\bea
\label{Vd2}
V =|H_2|^2\hat m^2(\alpha,\gamma)
+|H_2|^4\frac{1}{8}(g^2+g'^2)\left[1-\alpha^2-\gamma^2\right]^2
\eea
with
\be
\label{mgorro}
\hat m^2(\alpha,\gamma)=m_2^2+\left(m_{Q}^2 +m_{u}^2\right) \alpha^2
+m_1^2\gamma^2 - 2 m_3^2\gamma\;.
\ee

{\em If} $\alpha,\gamma$ are such that $\hat m^2(\alpha,\gamma)<0$, then
$V$ has a minimum in the $H_2$ direction at
\be
\label{H2mind}
|H_2|^2_{\rm min} = \frac{-4\hat m^2(\alpha,\gamma)}{(g^2+g'^2)
\left[1-\alpha^2-\gamma^2\right]^2}\;,
\ee
\be
\label{Vmind}
V_{\rm min}(\alpha,\gamma) = \frac{-2|\hat m(\alpha,\gamma)|^4}{(g^2+g'^2)
\left[1-\alpha^2-\gamma^2\right]^2}\;.
\ee
It is important to stress that this is {\em not} necessarily a CCB minimum
(in fact it will never be) since we have still to minimize with respect
to $\alpha,\gamma$ and we could well find $\alpha=0$ in that process.
Actually, the realistic minimum, $V_{\rm real\;min}$
(see eq.(\ref{Vreal})), is
a particular case of (\ref{Vmind}), more precisely
\be
\label{Vreald}
V_{\rm real\;min}=V_{\rm min}(\alpha=0,\gamma=\gamma_{\rm real})
= - \frac{ \left \{ \; \left[ \;( m_1^2+m_2^2 )^2-4 |m_3|^4 \; \right]
^{1/2}  - m_1^2+m_2^2 \; \right \} ^2  } {2 \; (g^2+{g'}^2) }
\ee
with
\be
\label{gammareal}
\left(\gamma_{\rm real}\right)^{-1}=\tan\beta=
\frac{m_1^2+m_2^2}{2m_3^2}+\sqrt{\left(
\frac{m_1^2+m_2^2}{2m_3^2}\right)^2-1}\;.
\ee
Of course, one has to demand
\be
\label{condd}
V_{\rm min}(\alpha,\gamma)> V_{\rm real\;min}
\;\;,
\ee
well understood that (\ref{condd}) does {\em not} necessarily mean
that we are comparing the relative depth of two minima of $V$, since
$V_{\rm min}(\alpha,\gamma)$ may not correspond to an actual minimum.

A necessary condition for (\ref{condd}) to be satisfied is that
\be
\label{derivd}
\left.\frac{\partial V_{\rm min}(\alpha,\gamma)}{\partial \alpha^2}
\right|_{\alpha=0,\gamma=\gamma_{\rm real}}
>0
\;\;.
\ee
This condition was worked out in ref.\cite{Gunion}, but without including
$\gamma$ in the game (the authors took $\gamma=0$). Now, it is clear that
(\ref{derivd}) corresponds to the requirement that $V_{\rm real\;min}$
is an actual minimum in the whole field-space (not just in $H_1,H_2$).
As it was mentioned in sect.2, this is simply equivalent to demand all
the scalar mass eigenvalues to be positive. If this is demanded from the
beginning (as it should be), eq.(\ref{derivd}) is a redundant condition.
In fact (\ref{derivd}) has the explicite form
\be
\label{mQmu}
m_Q^2+m_u^2>\frac{1}{2}\left|\left[ \;( m_1^2+m_2^2 )^2-4 |m_3|^4 \; \right]
^{1/2}  - m_1^2+m_2^2 \; \right|
\;\;,
\ee
which is equivalent to require that the sum of the two mass eigenvales of
the $u$--mass matrix is positive. For our later convenience, let us note
that (\ref{mQmu}) implies
\be
\label{mQmum2}
m_Q^2+m_u^2 + m_2^2>0
\;\;.
\ee

In order to study the relevance of (\ref{condd}) we must consider the
{\em minimum} of $V_{\rm min}(\alpha,\gamma)$ in the $\alpha,\gamma$
variables. It is interesting to check that $\alpha=0,\gamma=\gamma_{\rm real}$
does correspond to a minimum (the realistic one). However, there might
be other minima. A necessary condition to have a minimum is
$\hat m^2(\alpha,\gamma)<0$, which implies
\be
\label{mgamma0}
m_2^2+m_1^2\gamma^2-2m_3^2\gamma<0\;\;.
\ee
Using $m_1^2+m_2^2>2m_3^2$ (eq.(\ref{ufbhiggs})) and $m_1^2>m_2^2$,
it is clear that (\ref{mgamma0}) can only be satisfied in a certain
range of values of $\gamma$:
\be
\label{rangegamma}
0\le\gamma_{inf}\le\gamma\le\gamma_{sup}<1\;\;.
\ee
Now we can write the minimization condition for $\alpha$
\bea
\label{minalfa}
\frac{\partial V_{\rm min}(\alpha,\gamma)}{\partial \alpha^2}
&=&\frac{-4\hat m^2(\alpha,\gamma)}{
\left[1-\alpha^2-\gamma^2\right]^3}\left[
m_2^2+m_1^2\gamma^2-2m_3^2\gamma+(m_Q^2+m_u^2)(1-\gamma^2)
\right]
\nonumber \\
&=&\frac{-4\hat m^2(\alpha,\gamma)}
{\left[1-\alpha^2-\gamma^2\right]^3}\hat m^2(\alpha^2=1-\gamma^2,\gamma)
\;\;,
\eea
where the quantity $\hat m^2(\alpha^2=1-\gamma^2,\gamma)$ satisfies
\bea
\label{mhatcont}
\hat m^2(\alpha^2=1-\gamma^2,\gamma=0)&>&0
\nonumber \\
\hat m^2(\alpha^2=1-\gamma^2,\gamma=1)&>&0\;\;.
\eea
To analyze (\ref{minalfa}) we can distinguish two cases

\begin{description}

\item[a)] $m_1^2<m_Q^2+m_u^2$

In this case (which is the usual one)
$\hat m^2(\alpha^2=1-\gamma^2,\gamma)$ is a monotonically decreasing
function in the range $0\le\gamma\le 1$, so from (\ref{mhatcont})
it follows that
\be
\label{mmay}
\hat m^2(\alpha^2=1-\gamma^2,\gamma)>0
\ee
in all this range. Then, since
$\hat m^2(\alpha^2>1-\gamma^2,\gamma)>
\hat m^2(\alpha^2=1-\gamma^2,\gamma)$, it is clear
from the condition $\hat m^2(\alpha,\gamma)<0$
that (\ref{minalfa}) is only meaningful in a certain range
\be
\label{rangealfa}
0\le\alpha^2\le\alpha_{sup}^2<1-\gamma^2\;\;.
\ee
Hence, it follows from (\ref{mmay}), (\ref{minalfa}) that
$\frac{\partial V_{\rm min}(\alpha,\gamma)}{\partial \alpha^2}>0$
in all the range (\ref{rangealfa}), and therefore the optimum value
of $\alpha$ is always $\alpha=0$. Consequently, there are {\em no}
CCB minima.

\item[b)] $m_1^2>m_Q^2+m_u^2$

This is a rather unusual, but still possible case. Now
$\hat m^2(\alpha^2=1-\gamma^2,\gamma)$ is not monotonically decreasing
in the range $0\le\gamma\le 1$, but it has a minimum. However, if
\be
\label{extra}
m_2^2+m_Q^2+m_u^2 - \frac{m_3^4}{m_1^2-m_Q^2-m_u^2}>0
\;\;,
\ee
it is still true that $\hat m^2(\alpha^2=1-\gamma^2,\gamma)>0$ in all the
$0\le\gamma\le 1$ range. Then, the argument follows exactly as in the
previous case {\em (a)}. If (\ref{extra}) is not satisfied, then there is
a segment of $\gamma$ values where $\hat m^2(\alpha^2=1-\gamma^2,\gamma)<0$
(only the part of the segment overlaping (\ref{rangegamma}) is relevant).
For these values of $\gamma$ it is clear from (\ref{minalfa}) that
$\frac{\partial V_{\rm min}(\alpha,\gamma)}{\partial \alpha^2}<0$
for $\alpha^2<1-\gamma^2$ and
$\frac{\partial V_{\rm min}(\alpha,\gamma)}{\partial \alpha^2}>0$
for $\alpha^2>1-\gamma^2$. Therefore {\em there is} a CCB minimum
at $\alpha^2=1-\gamma^2$, {\em but} this is precisely the point where
D--terms$=0$. Note also from (\ref{Vmind}) that at this point
$V\rightarrow -\infty$, but this is not right since if D--terms$=0$,
we cannot neglect the terms involving $\lambda$ any more.

\end{description}

Finally, had we included the $SU(3)$ D--term in the game (relaxing
eq.(\ref{su3})), it is easy to convince yourself that the whole
argument would have followed analogously.

\subsection*{Property 5}

The last property concerns the optimum choice of the phases of the
fields involved in the scalar potential when analyzing CCB minima.
Taking again $\lambda Q H_2 u$ as the relevant coupling, the relevant
terms in the superpotential are
\be
\label{Wph}
W=\epsilon_{ij}
\lambda H_{2i} Q_j u +  \mu\epsilon_{ij} H_{1i} H_{2j}\;\; ,
\ee
The corresponding terms in the scalar potential without a definite phase,
say $V_{\rm ph}$, are
\bea
\label{Vph}
V_{\rm ph}&=&\left(A\lambda \epsilon_{ij}H_{2i} Q_j u + {\rm h.c.} \right)
\nonumber \\
&+&\left(B\mu\epsilon_{ij} H_{1i} H_{2j}+ {\rm h.c.} \right)
\ -\ \left(\mu^*\lambda H_{1i}^* Q_i u + {\rm h.c.} \right)
\eea
We will take $\lambda,\mu,A,B$ as real numbers for simplicity and also
because their phases are quite constrained by limits on the electric
dipole moment of the neutron since they give large one-loop contributions
to this CP-violating quantity.
The following results are independent of the signs of $\mu,\lambda$, as
well as on the form in which the two $SU(2)$ contractions in (\ref{Wph})
are defined. This comes from the fact that all these signs can be
re-absorved in phase redefinitions of the fields involved.
Along the direction $H_1^o, H_2^o, u_L, u_R \neq 0$ at which the CCB minima
appear (see text), $V_{\rm ph}$ can be re-writen as
\bea
\label{Vph2}
V_{\rm ph}&=&-2\left|A\lambda H_{2} Q u \right|
\ {\rm sign}(A)\ {\rm sign}(\lambda)\ \cos(\alpha+\beta)
\nonumber \\
&+&2\left|B\mu H_{1} H_{2}\right|
\ {\rm sign}(B)\ {\rm sign}(\mu)\ \cos(\alpha+\gamma)
\nonumber \\
&-&2\left|\mu\lambda H_{1}^* Q u \right|
\ {\rm sign}(\mu)\ {\rm sign}(\lambda)\ \cos(\beta-\gamma)
\;\;,
\eea
where $\alpha=$ phase($H_2^o$), $\beta=$ phase($u_Lu_R$),
$\gamma=$ phase($H_1^o$).

Of course, if $H_1=0$, the only non-vanishing term in (\ref{Vph2})
is the one proportional to $A$, which, for exploring minima of
the potential, can always be writen as
\bea
\label{VA}
V_{\rm ph}=-2\left|A\lambda H_{2} Q u \right|
\;\;.
\eea

If $H_1\neq 0$ and sign($A$) $=-$ sign($B$), it is straightforward to check
from (\ref{Vph2}) that $\alpha,\beta,\gamma$ can be taken so that the
three terms become negative, which of course corresponds to the deepest
direction in $V_{\rm ph}$, i.e.
\bea
\label{Vph3}
V_{\rm ph}&=&-2\left|A\lambda H_{2} Q u \right|
-2\left|B\mu H_{1} H_{2}\right|
-2\left|\mu\lambda H_{1}^* Q u \right|
\;\;.
\eea

If $H_1\neq 0$ and sign($A$) $=$ sign($B$), the previous direction
(\ref{Vph3}) is no longer available. Then $V_{\rm ph}$ can be expressed
as
\bea
\label{Vph4}
V_{\rm ph}=C_1\cos(\varphi_1)+C_2\cos(\varphi_2)+C_3\cos(\varphi_1-
\varphi_2)
\;\;,
\eea
where $C_i>0$ are the three absolute values of eq.(\ref{Vph2}), ordered
for convenience so that
\bea
\label{order}
C_1\ge C_2\ge C_3
\;\;,
\eea
and the $\varphi_i$ phases are certain independent combinations of
$\alpha,\beta,\gamma$ and the signs of $A,B,\lambda,\mu$.
For fixed values of $C_i$, the minimization in the $\varphi_1,\varphi_2$
variables gives the following result:

\begin{itemize}
\item If
\bea
\label{condC1}
\frac{C_2}{C_3}\ge 1+\frac{C_2}{C_1}
\eea
(this is by far the most usual case), then the minimum in the $\varphi_i$
space lies on
\bea
\label{minC1}
\varphi_1=\pi,\;\varphi_2=\pi
\;\;,
\eea
i.e. in this case $V_{\rm ph}$ can simply be expressed as
\bea
\label{Vph5}
V_{\rm ph}=-C_1-C_2+C_3
\;\;.
\eea

\item If
\bea
\label{condC2}
\frac{C_2}{C_3}\le 1+\frac{C_2}{C_1}
\;\;,
\eea
then the optimum choice of phases is given by
\bea
\label{minC2}
\left|\frac{\sin\varphi_2}{\sin\varphi_1}\right|&=&\frac{C_1}{C_2}
\nonumber \\
\left|\frac{\sin\varphi_1}{\sin(\varphi_1-\varphi_2)}\right|&=&\frac{C_3}{C_1}
\;\;,
\eea
which substituted in (\ref{Vph4}) gives
\bea
\label{minC22}
V_{\rm ph}=-\frac{1}{2}C_1C_2C_3\left(\frac{1}{C_1^2}+\frac{1}{C_2^2}+
\frac{1}{C_3^2}\right)
\;\;.
\eea

\end{itemize}

\noindent
Clearly, (\ref{condC2}) is
much more unlikely than (\ref{condC1}) and harder to handle
(compare eqs.(\ref{minC1},\ref{Vph5}) with eqs.(\ref{minC2}, \ref{minC22}).
Furthermore,
in the rare cases corresponding to (\ref{condC2}),
eqs.(\ref{minC1},\ref{Vph5}) still provide a very good
approximation\footnote{The worst situation occurs for
$C_1=C_2=C_3$, where the actual minimum of $V_{\rm ph}$ is $-3C_1/2$,
while eq.(\ref{Vph5}) gives $-C_1$.}
to the actual minimum of $V_{\rm ph}$. In consequence, we have always used
eq.(\ref{Vph5}) as the optimum direction of $V_{\rm ph}$ when
sign($A$) $=$ sign($B$).

\vspace{0.3cm}
\noindent
Finally, let us point out that all the previous results about the choice of
phases translate unchanged to the cases in which  the relevant coupling
is of the $\lambda Q H_1 d$ or $\lambda L H_1 e$ types.

\newpage

\def\MPL #1 #2 #3 {{\em Mod.~Phys.~Lett.}~{\bf#1}\ (#2) #3 }
\def\NPB #1 #2 #3 {{\em Nucl.~Phys.}~{\bf B#1}\ (#2) #3 }
\def\PLB #1 #2 #3 {{\em Phys.~Lett.}~{\bf B#1}\ (#2) #3 }
\def\PR #1 #2 #3 {{\em Phys.~Rep.}~{\bf#1}\ (#2) #3 }
\def\PRD #1 #2 #3 {{\em Phys.~Rev.}~{\bf D#1}\ (#2) #3 }
\def\PRL #1 #2 #3 {{\em Phys.~Rev.~Lett.}~{\bf#1}\ (#2) #3 }
\def\PTP #1 #2 #3 {{\em Prog.~Theor.~Phys.}~{\bf#1}\ (#2) #3 }
\def\RMP #1 #2 #3 {{\em Rev.~Mod.~Phys.}~{\bf#1}\ (#2) #3 }
\def\ZPC #1 #2 #3 {{\em Z.~Phys.}~{\bf C#1}\ (#2) #3 }

\newpage

\section*{Figure Captions}

\begin{description}
\item[Fig.~1] Excluded regions in the parameter space of the Minimal
Supersymmetric Standard Model, with $B=A-m$, $m=100$ GeV and
$M^{\rm phys}_{\rm top}=174$ GeV. The central darked region is excluded because
there is no solution for $\mu$ capable of producing the correct electroweak
breaking. The upper and lower darked regions are
excluded because it is not possible to reproduce the experimental mass of the
top. a) The circles and diamonds indicate regions excluded by the
``traditional"
Charge and Color Breaking constraints associated with
the $e$ and $d$-type trilinear terms respectively.
b) The same as (a) but using our ``improved" Charge and Color Breaking
constraints. The triangles correspond to the $u$-type trilinear terms.
c) The crosses, squares and small filled squares indicate
regions excluded by the Unbounded From Below-1,2,3 constraints respectively.
d) The previous excluded regions together with the one
arising from the experimental lower bounds on
supersymmetric particle masses (filled diamonds).

\item[Fig.~2] Excluded regions in the parameter space of the Minimal
Supersymmetric Standard Model, with $B=A-m$ and
$M^{\rm phys}_{\rm top}=174$ GeV, for
different values of $m$.
The central darked region is excluded because
there is no solution for $\mu$ capable of producing the correct electroweak
breaking. The upper and lower darked regions are
excluded because it is not possible to reproduce the experimental mass of the
top. The small filled squares indicate regions excluded by our
Unbounded From Below constraints. The circles indicate regions excluded
by our ``improved" Charge and Color Breaking constraints. The filled diamonds
indicate regions excluded by the experimental lower bounds on supersymmetric
particle masses. The ants indicate regions excluded by
negative scalar squared mass eigenvalues.

\item[Fig.~3] The same as Fig.~2 but with $B=2m$. Now, the whole
darked region is excluded because it is not
possible to reproduce the experimental mass of the top.

\item[Fig.~4] Contours of allowed regions in the parameter space of the Minimal
Supersymmetric Standard Model, with
$M^{\rm phys}_{\rm top}=174$ GeV  and different
values of $B$ and $m$, by the whole set of constraints.

\end{description}

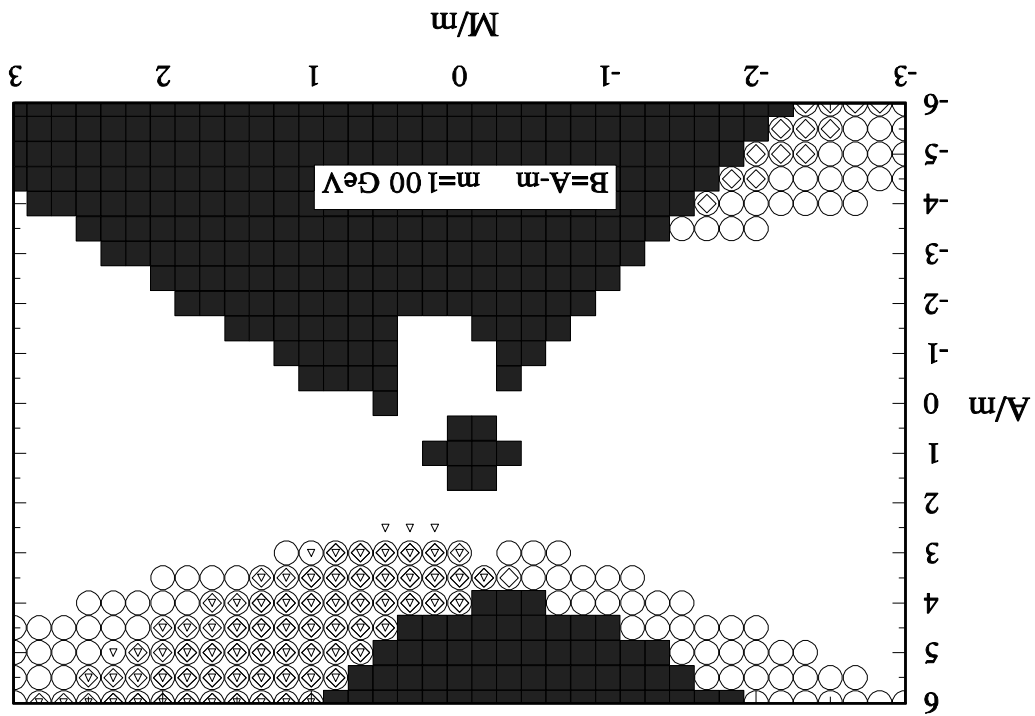
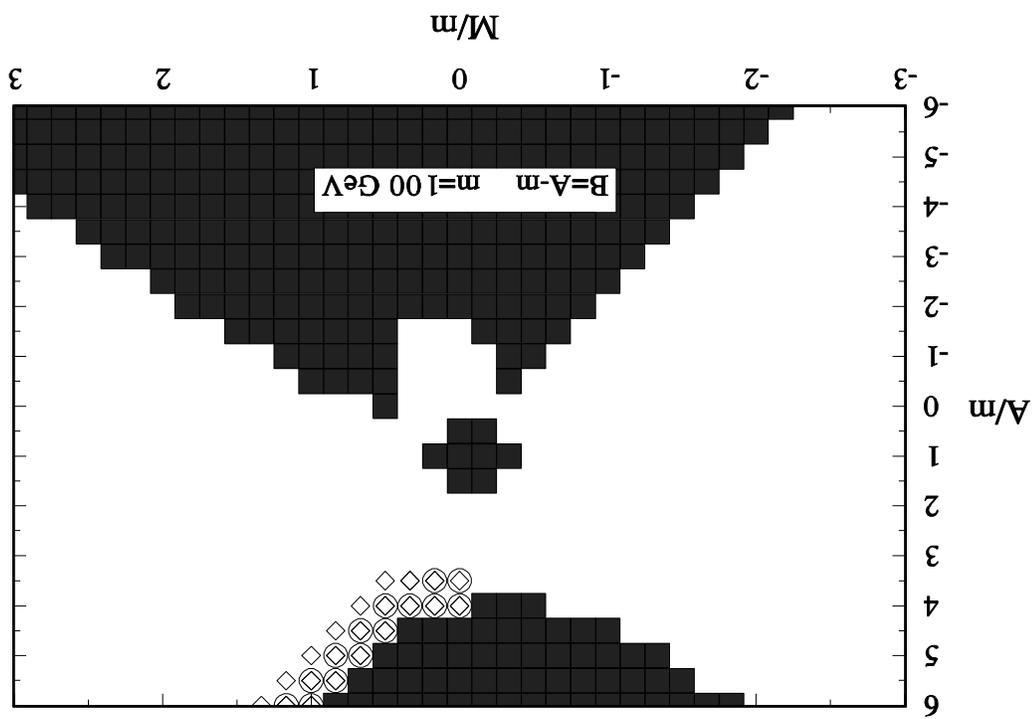
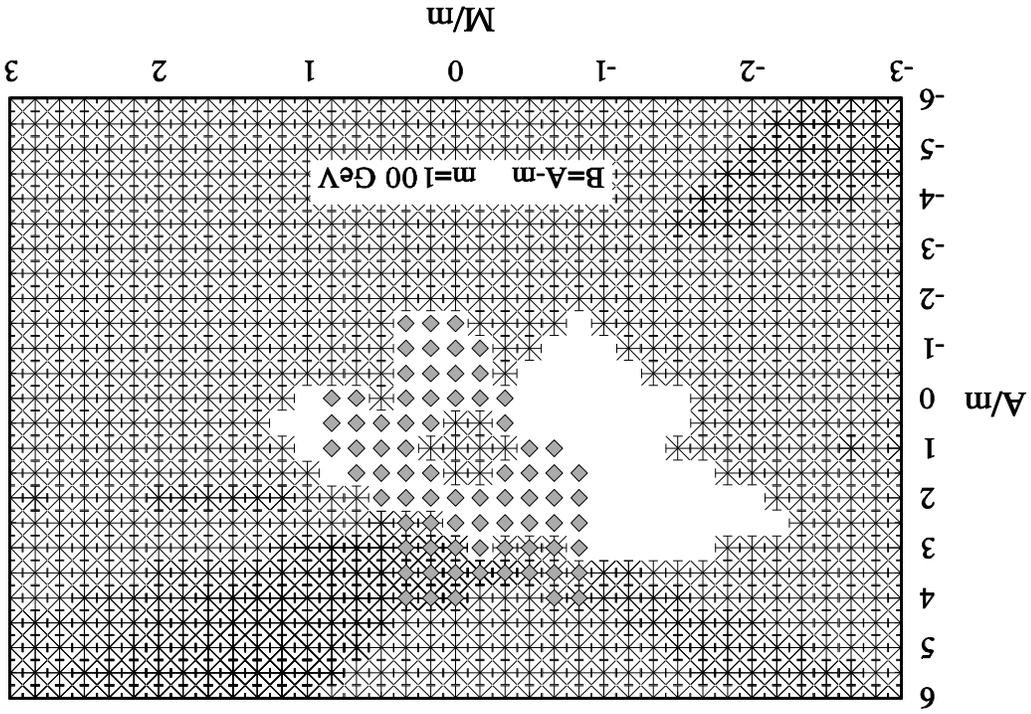
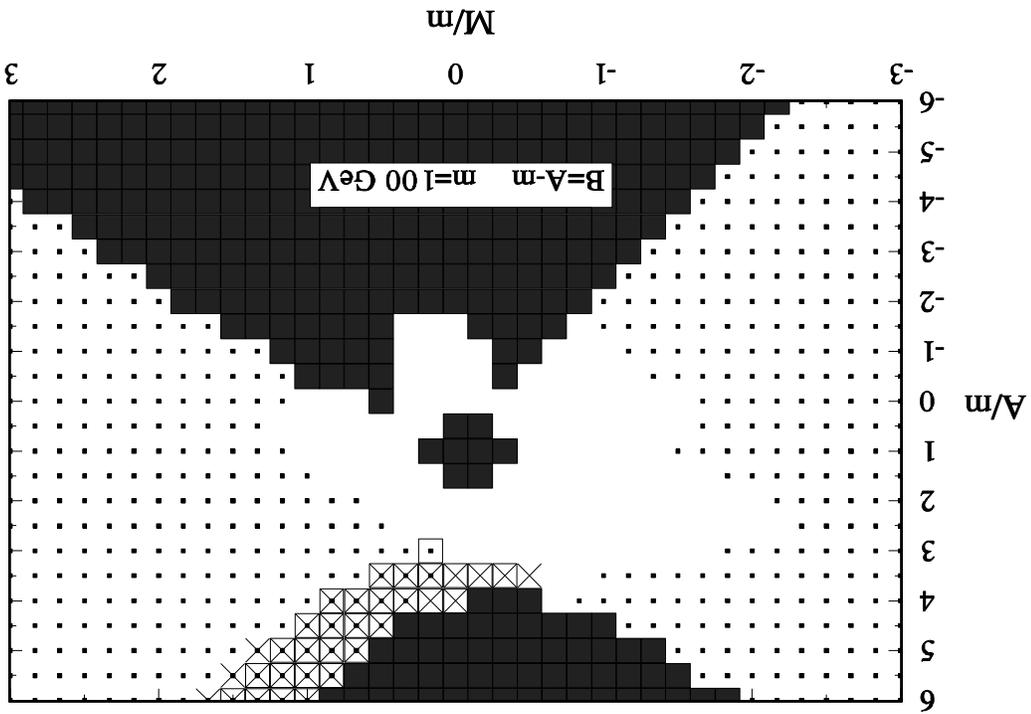
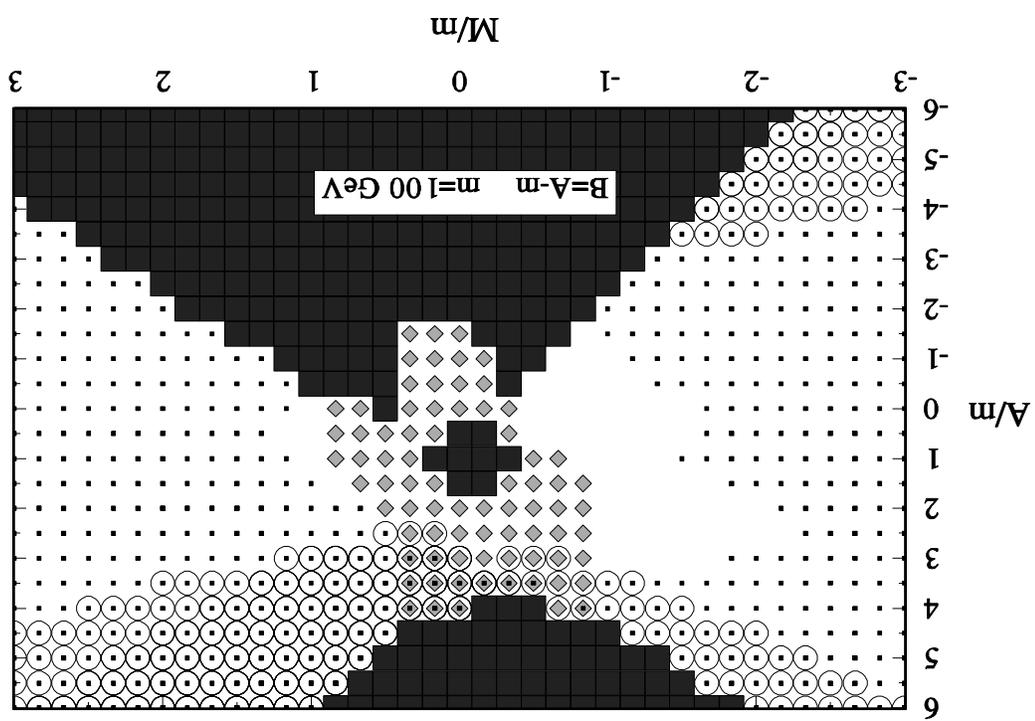
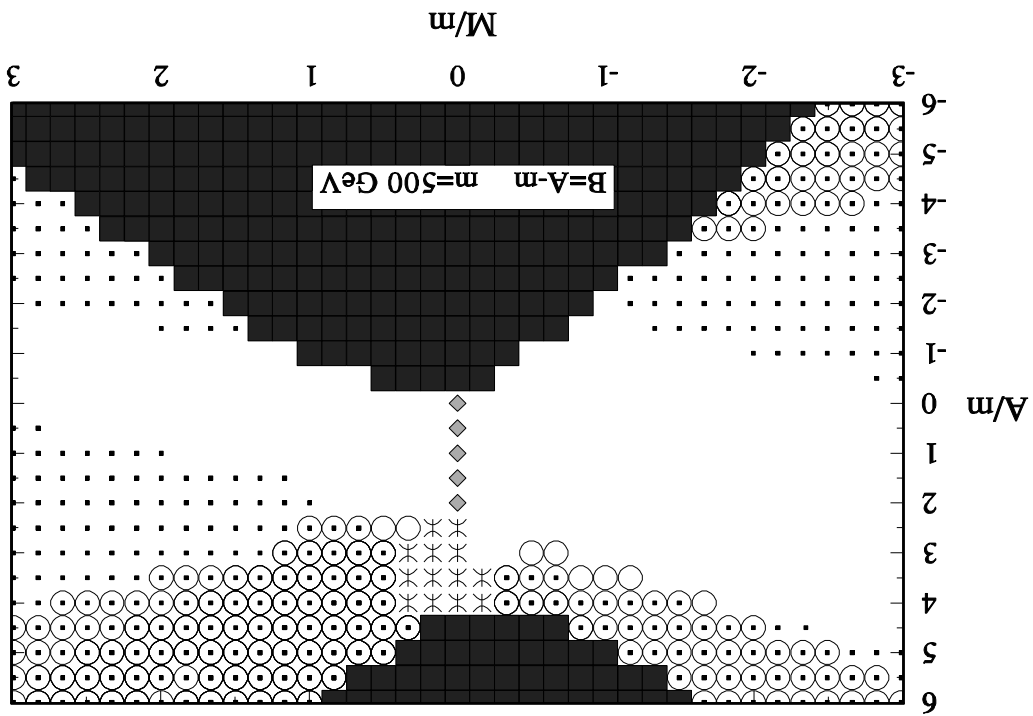
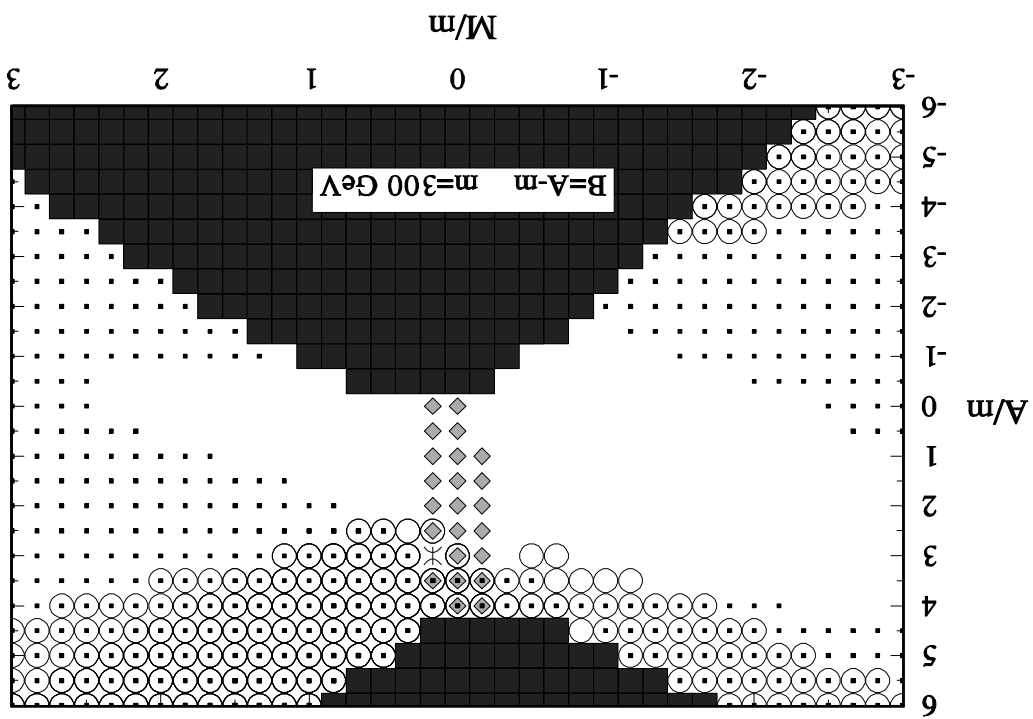
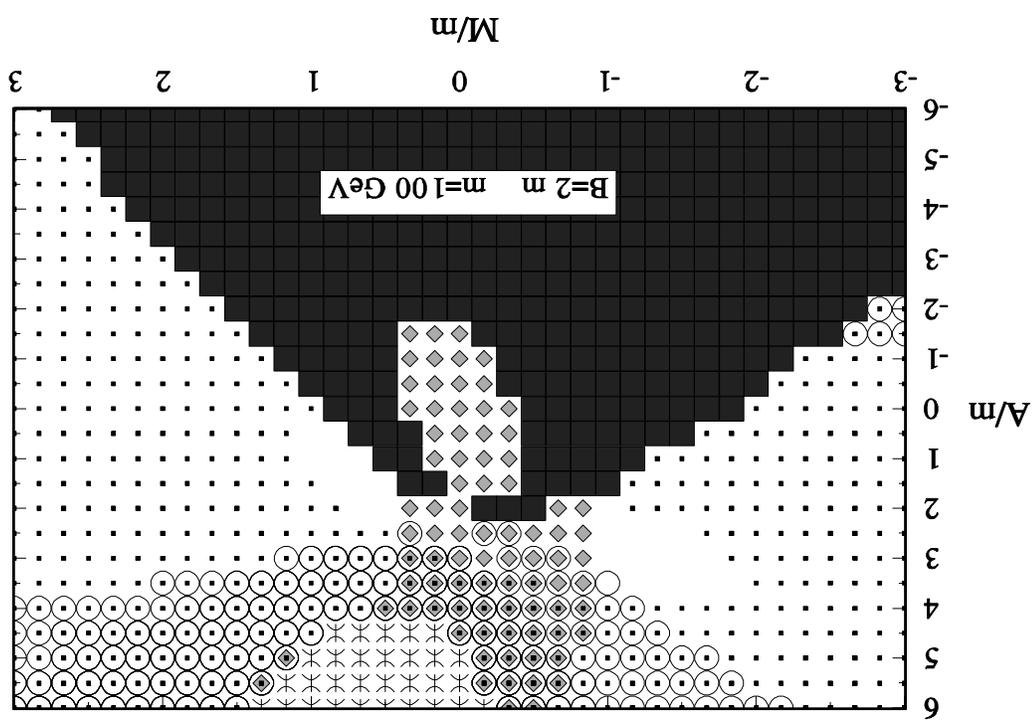
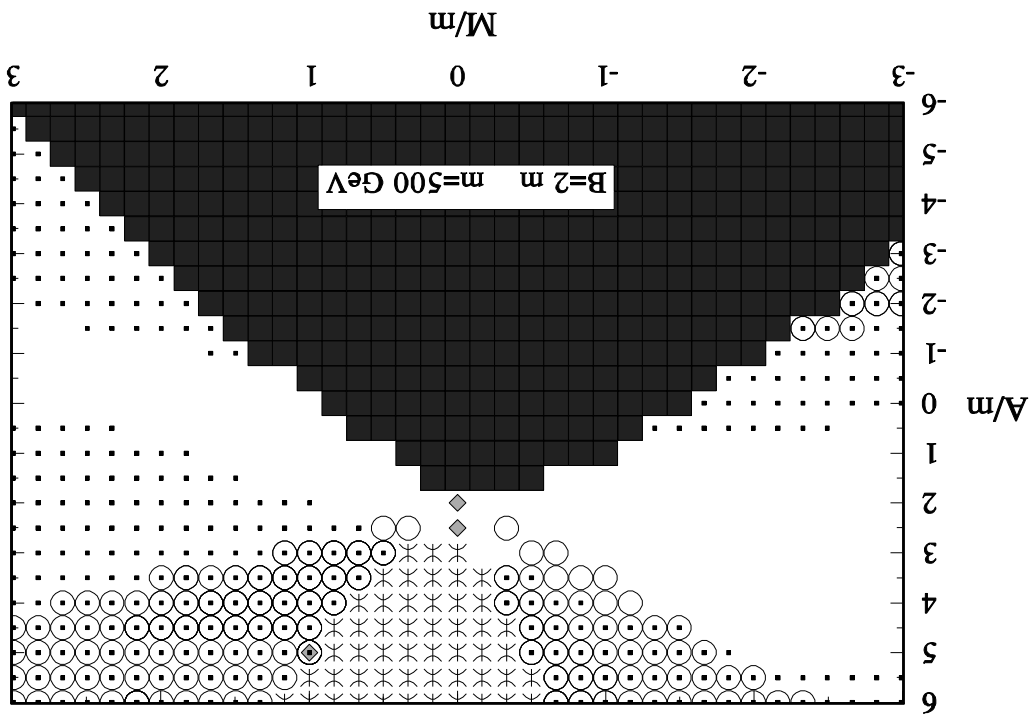
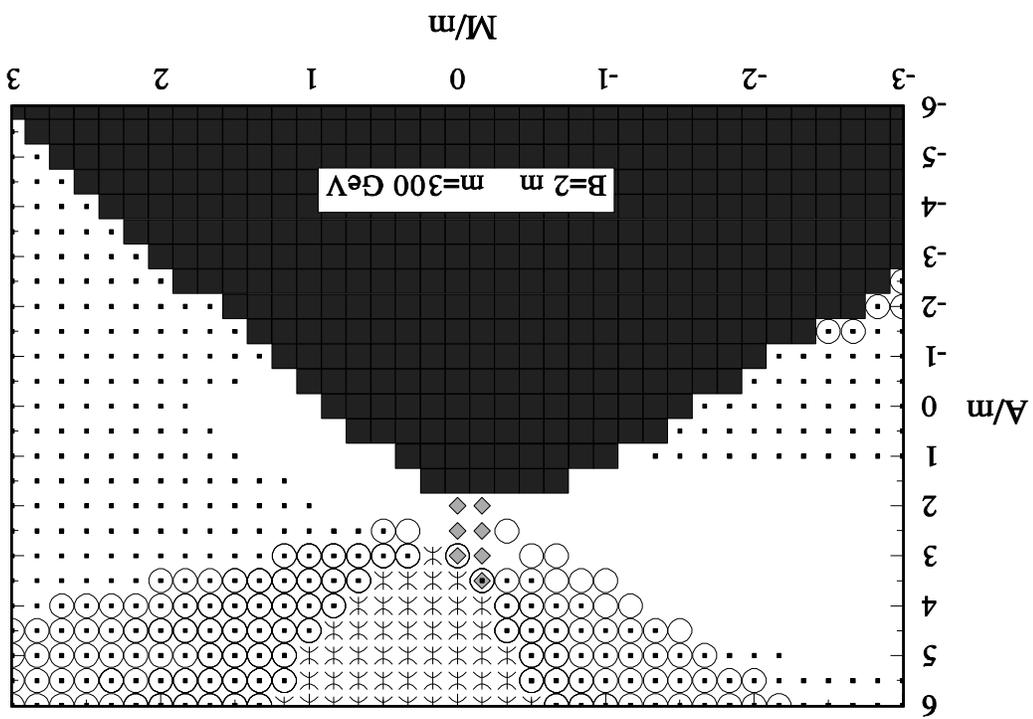
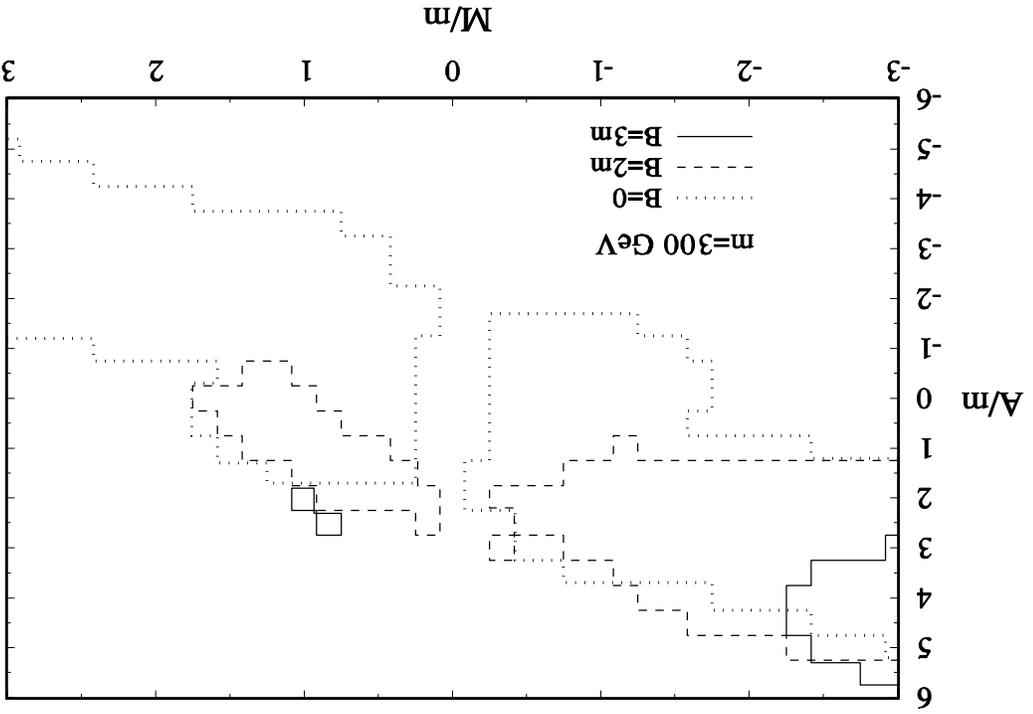
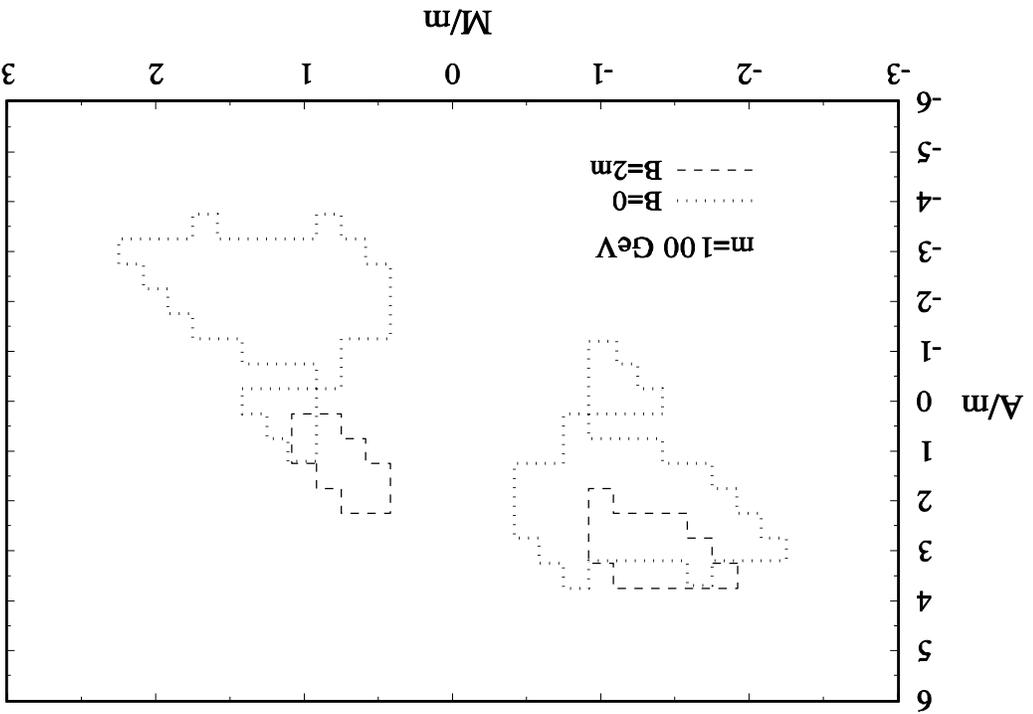
\end{document}